\documentclass[twocolumn,aps,showpacs,pra]{revtex4}
\usepackage{amssymb}
\usepackage{mathrsfs}
\usepackage{amsfonts}
\usepackage{graphicx}%
\usepackage{dcolumn}
\usepackage{amsmath}

\newcommand{\be}{\begin{equation}}
\newcommand{\ee}{\end{equation}}
\newcommand{\bey}{\begin{eqnarray}}
\newcommand{\eey}{\end{eqnarray}}
\newcommand{\ba}{\begin{array}}
\newcommand{\ea}{\end{array}}
\newcommand{\bi}{\begin{itemize}}
\newcommand{\ei}{\end{itemize}}
\newcommand{\bem}{\begin{enumerate}}
\newcommand{\eem}{\end{enumerate}}
\newcommand{\bw}{\begin{widetext}}
\newcommand{\ew}{\end{widetext}}
\newcommand{\ra}{\rangle}
\newcommand{\la}{\langle}
\newcommand{\pp}{\partial}
\newcommand{\ov}{\overline}

\newcommand{\ww}{\widetilde}

\newcommand{\E}{{\cal E}}

\newcommand{\HH}{{\mathscr{H}}}

\newcommand{\PP}{{\mathcal{P}}}
\newcommand{\D}{{\mathcal{D}}}
\newcommand{\Pm}{{\mathcal{P}_{\mu}}}
\newcommand{\Pnn}{{\mathcal{P}_{\nu}}}
\newcommand{\Pmp}{{\mathcal{P}_{\mu '}}}
\newcommand{\R}{{\mathcal{R}}}

\newcommand{\cs}{\mathcal{S}}

\newcommand{\tr}{{\rm Tr}}

\begin{document}

 \title{A quantum theory for a total system including a reference system}

\author{Wen-ge Wang}
\affiliation{
 Department of Modern Physics, University of Science and Technology of China,
 Hefei 230026, China
 }

 \date{\today}

 \begin{abstract}

 The standard formalism of quantum mechanics is extended to describe a
 total system including the reference system (RS), with respect to which the total
 system is described.
 The RS is assumed to be able to act as a measuring apparatus,
 with measurement records given by the values of some reference properties of the RS.
 In order to describe the total system, we define
 a frame of reference (FR) as a set of states that can be used to express
 all other states of the total system.
 The theory is based on four basic postulates, which have, loosely speaking,
 the following contents.
 (i) A reference property of a RS has a definite value and is sufficiently stable
 in the FR directly related to the reference property.
 (ii) States of the total system are associated with vectors in the Hilbert space.
 (iii) Schr\"{o}dinger equation is the dynamical law in each valid FR.
 (iv) Under certain condition a property of a system can be regarded as
 a reference property; vector descriptions of the total system given in different
 FRs of the same RS may have a probabilistic relationship like in Born's rule.

 As a result of the four postulates, the same state of the total system in the same FR
 may have multiple descriptions, some given by pure vectors and some given by density
 operators. For the consistency of the descriptions, a principle is introduced,
 which states that the descriptions must be physically equivalent in the sense that
 they give the same predictions for measurement results. As an important consequence
 of the principle of consistent description, it imposes a restriction to vectors
 in the Hilbert space that can be associated with physical states.
 This restriction breaks the time reversal symmetry: The time reversal vector of
 a physically allowed vector may be physically not allowed.
 In particular, von Neumann entropy for the total system is found to be able to
 keep constant or increase with time, but never decrease with time.
 Finally, we give an example to illustrate the theory, in which a two-level system is taken
 as a reference system
 and compare the proposed theory with many-worlds
 interpretations and consistent-histories interpretations of quantum mechanics.

 \end{abstract}

 \maketitle


 \section{Introduction}

 In the standard formalism of quantum mechanics (see textbooks, e.g., \cite{Neumann-qm}),
 measurement is done by some outside measuring
 apparatus, hence, processes involving and not involving measurement are assumed to have
 different types of time evolution:
 The state vector of a system has continuous Schr\"{o}dinger evolution when no measurement is performed,
 while has discontinuous change in a measuring process,
 namely, the so-called wave packet reduction or collapse of state vector.

 When extending the standard formalism of quantum mechanics to a theory
 including the measuring apparatus,
 one faces the problem of potential confliction between Schr\"{o}dinger evolution and
 definite outcomes of measurement.
 This gives rise to the so-called measurement problem, a topic of
 debating since the establishment of quantum mechanics.
 To solve this problem, there have been numerous efforts,
 introducing various types of interpretation of quantum mechanics
 (see, e.g., reviews given in Refs.~\cite{Laloe01,Zurek03,JZKGKS03,BG03,Schloss04}),
 but, none of them has been commonly accepted.


 In this paper, we study a direct strategy of extending the standard formalism of quantum
 mechanics to include the measuring apparatus:
 We take the definiteness of measurement records as a basic assumption,
 then, consider how to adjust the standard formalism of quantum mechanics to form a
 extended formalism.
 An important reason of introducing this assumption is that it has sound experimental basis.
 A purpose of adopting this strategy is to see which type of formalism it will lead to,
 if both the definiteness of measurement records and Schr\"{o}dinger equation are
 assumed at the fundamental level.

 We would emphasize on the unique position that measuring apparatus occupies
 in our understanding of the world:
 (1) Its record is the source of all available information about the world,
 and in this sense it is the original ``describer'';
 (2) it participates in interaction, hence, is an participant;
 (3) its records gives the ultimate judgement on predictions of all theories,
 and in this sense it acts like a ``judge''.
 Fulfilling the three requirements is not an easy task for a quantum theory.

 One of our observations, which has not been discussed in the literature, is
 that a measuring apparatus can also play the role of a reference system (RS),
 hence, the concept of RS should be addressed at the fundamental conceptual level of the theory.
 Indeed, some changes in the conceptual structure are necessary in such an extension of
 the standard formalism of quantum mechanics and this is discussed in Sec.~\ref{sect-ana-con}.
 In particular, since the measuring apparatus is to be described within the theory,
 measuring processes are also to be described, hence, measurement can not be a basic concept
 in the proposed theory.

 Basic assumptions are to be introduced in Sec.~\ref{sect-basic-p}.
 We keep the following contents of the standard formalism of quantum mechanics:
 namely, the Hilbert space for describing physical states, Schr\"{o}dinger equation
 for time evolution, and the essential part of Born's rule.
 Meanwhile, we add the following new elements to the formalism: namely,
 (i) association of a frame of reference (FR) with some reference property of a RS,
 (ii) the definiteness and stableness of a reference property of a RS
 in its own FR, such that the RS may act as a measuring apparatus
 with the value of that property serving as measurement record,
 and (iii) a criterion for a FR to be valid.
 In addition, we introduce a principle of physical description, which states that
 descriptions given in the same FR for the same state must give
 physically equivalent predictions.

 More detailed discussions in physically equivalent description are given
 in Sec.~\ref{sect-BC}.
 In particular, it is shown that, under certain condition,
 a pure vector description may be physically equivalent to some density operator
 descriptions for the same state of the total system, in the sense that
 they give the same predictions for measurement results.
 In Sec.~\ref{sect-CC}, we derive an explicit expression for the principle of
 consistent description, which imposes a restriction to vectors in the
 Hilbert space that can be associated with physical states.

 Some applications of the theory are discussed in Sec.~\ref{sect-app},
 including a derivation of the contents of the axiom of measurement in the standard formalism
 of quantum mechanics.
 It is shown that von Neumann entropy of the total system may keep constant or increase,
 but never decrease, hence, time evolution of the total system is irreversible.
 Those vectors in the Hilbert space, for which von Neumann entropy decreases,
 are physically forbidden by the principle of consistent description.
 In Sec.~\ref{sect-model}, we discuss a simple model,
 in which a two-level system can server as a RS.
 Comparisons of the proposed theory with some interpretations of
 quantum mechanics are given in Sec.~\ref{sect-ot-int}.
 Finally, we give conclusions and discussions in Sec.~\ref{sect-conclusion}.

 \section{Conceptual structure}
 \label{sect-ana-con}

 In this section, we discuss the basic conceptual structure of the proposed theory,
 first giving some preliminary analyses, then, introducing the basic concepts.

 \subsection{Preliminary analysis (I): for some concepts in the standard formalism}

 As discussed above, in a quantum theory for a total system including the measuring apparatus,
 \emph{measurement should not be a basic concept}.
 Another concept that needs careful consideration is observable.
 Let us examine a typical measuring process,
 through which an observer can obtain information about some observable of a system.
 For this purpose, the observer needs a measuring apparatus,
 which may have some definite and stable properties that can serve as measurement records.
 The observer takes record of the definite properties,
 by directly reading or by reading after some amplification process,
 then, uses some theory to explain the measurement records and
 gets the measured value of the observable of the system.

 The above analysis shows that, in a theory describing both the measured system
 and the measuring apparatus,
 {observable of the measured system should not be a fundamental concept} and
 \emph{the most directly measured quantity is some recordable property of the
 measuring apparatus}.
 Such recordable properties are in fact the source of all information that one may
 obtain from measurement.

 It is also important to address the concept of state.
 A basic requirement for the existence of a physical description of a system
 is that the description has an objective feature in the sense that it
 does not depend on any specific feature or property of the observer,
 that is, it does not matter who performs the measurement and does the explanation.
 To express this point, we usually say that the system has a \emph{state}
 and the description is a description of the state of the system.
 {State should be a basic concept in each physical theory}.

 Due to the objective feature of state, when discussing the state of a system,
 it is not necessary to explicitly mention the observer,
 just keeping in mind that such an observer may exist in principle \cite{note-observer}.
 Moreover, since the mere function of an amplification process is to make it possible for
 the observer to read measurement records, such processes are not essential in the study
 here and will not be considered further.

 \subsection{Preliminary analysis (II): for concepts related to
 reference system}
 \label{sect-concept-II}

 As discussed above, measured properties of a measured system are derived from
 measurement records that correspond to some definite recordable properties
 of the measuring apparatus.
 This implies that the measuring apparatus also plays the role of a RS.
 Indeed, as well known, there exists no absolute state of a system
 and a state of a system must be described with respect to some RS.

 Since measurement record is the source of information obtainable from measurement,
 the state of a measured system should be a description with respect to
 the \emph{accessible recordable properties} of the measuring apparatus,
 not with respect to other properties of the apparatus.
 For this reason, we call an accessible recordable property of a measuring apparatus
 a \emph{reference property} of the measuring apparatus.
 The essential feature of a reference property is its definiteness;
 meanwhile, it should be stable enough, such that there is enough time for
 its value to be recorded.
 For example, in a Stern-Gerlach experiment, in which a particle hits a screen
 and leaves some record on the screen by inducing some chemical change to
 the molecular structure of the screen, the stable molecular structure of the screen
 can be regarded as a reference property of the screen.

 The existence of a RS is necessary for a description, but does not guarantee it.
 In fact, a description of a state of a system is given in terms of some other states,
 more exactly, by means of its relation to some other states.
 For example, in classical mechanics, one may choose a coordinate system
 and express the position of a particle in terms of its components in the coordinates.
 The situation is similar in the usual quantum mechanics:
 When the state of a quantum system is written as $|\Psi\ra = \sum_i c_i|i\ra $,
 it is expressed in terms of some other states associated with vectors $|i\ra $
 in the Hilbert space.

 Usually, with respect to a fixed RS,
 one may choose a collection of states to form a framework for description,
 such that all other states can be described in terms of these states.
 Such a collection of states can be called a ``frame of reference'' (FR),
 as a generalization of the usual usage of the concept, which refers to a coordinate
 system \cite{note-RF}.

 \subsection{Concepts}
 \label{sect-concept}

 Based on the analyses given above, now we introduce the conceptual structure of the theory.
 First of all, we introduce two most fundamental concepts.
 One is physical \emph{system}, which we use to indicate something that really exists.
 The other is the \emph{state} of a physical system, which
 we use to indicate the status of a system.

 Next, we consider the concepts of measuring apparatus, RS,  and reference property.
 As discussed above, these are closely related concepts.
 In fact, as shown below, we can use the concept of \emph{reference property}, to
 define the other two concepts.
 We stress again that the concept of reference property has its roots in the observation of
 the existence of definite recordable properties of measuring apparatuses in all experiments.


 We definite A \emph{reference system } (RS) as a physical system that possesses
 a reference property.
 A RS may have two reference properties at the same time.
 But, usually it is not necessary to stress this point, because due to the definiteness of
 reference properties, two reference properties may be combined to
 form a new reference property.
 We use $\R$ to indicate a  RS.
 The whole of all other systems of relevance to $\R$
 is called the environment of $\R$, denoted by $\E$.
 We emphasize that in this paper we consider this definition of RS for the purpose
 of developing a quantum theory for a total system including the measuring apparatus.
 We do not imply that this is the most general definition of RS.

 A \emph{measurement} is defined as
 a process of interaction between a RS $\R$ and some part of its
 environment $\E$, in which some reference property of $\R$ may change.
 A RS $\R$ defined above can in fact play the role of a measuring apparatus.
 To see this point, let us consider a process of interaction between $\R$ and a system $\cs$,
 and assume that the interaction is capable of changing a reference property of $\R$.
 Recording the values of the reference property of $\R$ before and
 after the interaction process,
 one may use the records to obtain information about the system $\cs$.
 This is essentially what we mean for a measurement process.

 A reference property of a system is a property of the state of the system,
 hence, it is not a basic concept.
 However, at the present stage we can not give a definition for reference property,
 since basic assumptions for properties of systems have not been introduced.
 The definition of reference property will be given in a later section,
 when certain basic postulates are introduced.

 We use the name \emph{frame of reference} (FR) to indicate a collection of physical states
 of a total system $\R +\E$, with respect to the RS $\R$ (more exactly, with respect to
 the employed reference property of $\R$),
 such that all other physical states of the total system can be expressed
 in terms of the states in this collection.
 Like the fact that there exist infinitely many coordinate
 systems for the same three-dimensional real space,
 there exist infinitely many FRs that are equivalent in the sense of
 giving equivalent descriptions of possible states of $\R+\E$.
 In a quantum theory, a FR is associated with a set of basis vectors in the Hilbert space.
 In what follows, for brevity, when speaking of a FR, sometimes we mean this FR together
 with all its equivalent FRs if no confusion is induced.


 \section{Basic assumptions}
 \label{sect-basic-p}

 In this section, we introduce the basic assumptions: four postulates and
 a principle for consistent description.
 We also discuss some features of reference property.

 \subsection{Postulates  I--III}
 \label{sect-I-III}

 In the first postulate, we assume the relation between RS and FR.
 \bi
 \item \textbf{Postulate I}:
 Related to each reference property of a RS, there exists a FR, and
 this reference property has a definite value and is sufficiently stable in this FR.
 \ei
 Here, the stableness of a reference property at a time $t$ means that the property has the
 same definite value at least within a time scale $\tau_s$ under dynamical evolution,
 i.e., for the time interval $[t,t+\tau_s]$.
 This requirement comes from the fact that a finite time interval is needed
 for a record to be taken.
 The time scale $\tau_s$ may be system-dependent.
 (In Sec.~\ref{cond-decoh}, we show that it is reasonable to require that $\tau_s$
 is larger than a decoherence time.)

 \emph{Some remarks}:
 (i) The assumption of the existence of a FR for each reference property of a RS
 is related to the fact that, as discussed in Sec.~\ref{sect-concept-II},
 a description of a measured system is obtained by
 explaining the value(s) of a reference property of the measuring apparatus.
 (ii) Here, we assume the definiteness of a reference property of a RS in its own FR,
 not in an arbitrary FR.


 In the second postulate, we assume the mathematical structure for
 states of the total system.
 \bi
 \item \textbf{Postulate II}:
 In each FR of a RS $\R$, the state of the total system $\R+\E$
 can be associated with a vector or a density operator in the total Hilbert space $\HH$.
 \ei
 We use $\HH_\R$, $\HH_\E$, and $\HH$ to denote the Hilbert spaces corresponding to a RS $\R$,
 its environment $\E$, and the total system $\R+\E$, respectively,
 with $\HH = \HH_{\R}\otimes \HH_{\E}$.
 When a state is described by a density operator $\rho$, we give it an ensemble interpretation.

 We remark that the extension of the Hilbert space considered in the usual
 quantum mechanics, which does not include the RS, to the total Hilbert space in
 Postulate II is non-trivial.
 In fact, that a RS can be described in its own FR is an assumption.
 This assumption allows a unified description for a RS and its environment.

 In the third postulate, we assume Schr\"{o}dinger equation as the dynamical law.
 \bi
 \item \textbf{Postulate III}:
 In each valid FR of $\R$, when the state of the total system $\R+\E$
 is described by a vector $|\Psi(t)\ra$ in the total Hilbert space $\HH$,
 its time evolution obeys Schr\"{o}dinger equation
 \be i\hbar \frac{\pp}{\pp t}|\Psi(t)\ra = H |\Psi(t)\ra . \label{evolve-RI}  \ee
 \ei
 Here, we use $H$ to denote the Hamiltonian of the total system $\R+\E$,
 \be H=H_{\R} + H_{\E} + H_I, \ee
 where $H_{\R}$ and $H_{\E}$ are the Hamiltonians of $\R$ and $\E$, respectively,
 and $H_I$ is the interaction Hamiltonian for the interaction between $\R$ and $\E$.
 We use $U(t,t_0)$ to denote the unitary evolution operator,
 \be |\Psi(t)\ra = U(t,t_0) |\Psi(t_0)\ra . \label{U} \ee
 In the case of a time-independent Hamiltonian, $U(t,t_0)= e^{-iH(t-t_0)/\hbar}$.
 As a consequence of Postulate III, the time evolution of a density operator $\rho$
 is given by
 \be \rho (t) = U(t,t_0)\rho(t_0) U^\dag (t,t_0).  \ee


 \subsection{Reference property}
\label{sect-RP}

 In this section, making use of the three postulates introduced above, we give further discussions
 for the concept of reference property and propose a definition for it.
 As a recordable property, a reference property may take discrete values.
 Generally, a property of $\R$ with discrete values $\mu = \mu_1, \mu_2, \ldots $
 can be associated with some division of $\HH_\R$, the Hilbert space of $\R$,
 into sub-regions denoted by $\HH_{\R\mu}$.

 In an ideal case of reference property, the corresponding sub-regions $\HH_{\R\mu}$
 are orthogonal subspaces of $\HH_\R$.
 In this paper, we consider this ideal case.
 In this case, the subspaces $\HH_{\R\mu}$ correspond to a division of $\HH_\R$,
 which is given by a set of orthogonal and complete projection operators $\Pm$,
 denoted by $\{\Pm\}$, satisfying
 \be \Pm\PP_{\nu} = \delta_{mn} \Pm , \ \ \ \  \sum_{\mu} \Pm = I_{\R}, \label{uni-R} \ee
 where $I_{\R}$ is the identity operator in the Hilbert space $\HH_{\R}$.
 Correspondingly, the total Hilbert space is also divided into a series of sub-spaces
 \be \HH_{\mu } \equiv \HH_{\R \mu }\otimes \HH_{\E} . \ee
 Obviously, $\Pm \otimes I_{\E}$ is the projection operator for the subspace $\HH_{\mu}$,
 where $I_{\E}$ is the identity operator in $\HH_{\E}$.
 For brevity, without the risk of confusion we also use $\Pm$ to indicate $\Pm \otimes I_{\E}$.

 We also use $\{\Pm\}$ to indicate the division of $\HH_\R$ (equivalently $\HH$) into
 the corresponding subspaces.
 And, we use $\R_{\{\mu\}}$ to indicate a RS $\R$ with a reference property
 related to $\{\Pm \}$.
 These abbreviation may bring much convenience.
 For example, when speaking of a FR that is related to a reference property of $\R$
 given by the division $\{\Pm \}$, we can simply say the FR of $\R_{\{\mu\}}$.
 Thus, according to Postulate I, when a FR of $\R_{\{\mu\}}$ is valid,
 the state of the total system $\R+\E$ in this FR has a definite value $\mu$ of the reference
 property related to $\{\Pm\}$.

 Obviously, a necessary condition for a property to be a reference property is that making
 this assumption does not lead to confliction between Postulates I and III.
 We assume that this necessary condition is also a sufficient one.
 This gives a definition for reference property.
 \bi
 \item
 A system $\R$ may have a \emph{reference property} related to a division $\{\Pm\}$
 for a vector description $|\Psi\ra \in \HH_\mu$ of the total system in the FR of
 $\R_{\{\mu\}}$ at a time $t$, if assuming this
 does not lead to confliction between Postulates I and III for this time.
 \ei
 Here, no confliction means that the assumed reference property of $\R$ may keep its
 definite value within a time period $\tau_s$ under Schr\"{o}dinger evolution.

 As an example of reference property,
 let us consider a trivial division of $\HH_\R$, denoted by $\{I\}$,
 which is given by the identity operator $I$ as the projection operator.
 This trivial division gives in fact no real division.
 Without the risk of inducing confusion, we also use $\{I\}$ to indicate
 the property of $\R$ related to the trivial division.
 Intuitively, the trivial property $\{I\}$ of a system just means the existence of the system.
 Clearly, the property $\{I\}$ is always definite and stable, since it may take one value only.
 Therefore, according to the definition of reference property, the trivial property
 $\{I\}$ of each system can always be taken as a reference property.
 As a result, the FR of $\R_I$ is always valid for an arbitrary system $\R$
 and the total system can always be described in this FR.
 Here and hereafter, for brevity, we use $\R_I$ to denote $\R_{\{I\}}$.



 \subsection{Postulates IV}
 \label{sect-Postulate-IV}

 In principle, a same state of the total system may be described in different FRs of $\R$.
 In the last postulate, we consider the relation between descriptions given in different FRs
 of the same RS.
 (In this paper, we do not consider the relation between FRs of different
 RS and would leave this to future investigation.)

 For example, let us consider a state of the total system that is described by $|\Psi\ra $
 in the FR of $\R_I$, which is always valid.
 The question here is whether a non-trivial FR of $\R_{\{\mu\}}$ is valid and, if valid,
 what is the description of the total system in this non-trivial FR.
 If $|\Psi\ra $ lies in one subspace $\HH_\mu$, we can directly use the definition of
 reference property to determine whether the property related to the division $\{\Pm\}$
 can be a reference property.
 If the property can be a reference property, then, the FR of $\R_{\{\mu\}}$ is valid and
 the total system is described by the same vector $|\Psi\ra $ in this new FR.

 More subtle is the case that $|\Psi\ra $ lies in more than one subspaces $\HH_\mu$.
 In this case, $|\Psi\ra $ can not be a description in the FR of $\R_{\{\mu\}}$,
 since this would conflict with Postulate I.
 If we assume that the FR of $\R_{\{\mu\}}$ is absolutely invalid in this situation,
 we may have a complete formalism of a theory.
 However, such a theory is not what we intend to propose, because it can not describe
 measuring processes.
 To see this point, let us consider a measuring process, in which the interaction between
 $\R$ and the measured system induces non-zero coupling between different
 subspaces $\HH_\mu$, and after which the measurement record takes the form a definite
 value of $\mu$.
 In this measuring process, Schr\"{o}dinger evolution leads to
 a $|\Psi\ra $, which is a superposition of components in more than one subspaces $\HH_\mu$.
 If a reference property related to $\{\Pm\}$ is absolutely impossible for this $|\Psi\ra $,
 then, the theory can not predict the possibility of taking a measurement record
 as a definite value of $\mu$.
 In fact, in such a theory, as a result of Schr\"{o}dinger evolution, the trivial reference
 property $\{I\}$ is usually the only reference property of $\R$, which implies that usually
 no measurement can be done.
 Therefore, in order to explain measurement, under certain condition,
 $|\Psi\ra $ should allow the possibility of some non-trivial reference property.

 Let us consider what may happen if a FR of $\R_{\{\mu\}}$ is valid.
 In this case, according to Postulate I,
 the state of $\R+\E$ in the FR of $\R_{\{\mu\}}$ should have a definite value of $\mu$.
 The simplest and most natural assumption for the state of the total
 system in the FR of $\R_{\{\mu\}}$ would be one of $\Pm|\Psi\ra $,
 if the condition stated in the definition of reference property is satisfied.
 Since $|\Psi\ra $ has more than one components $\Pm|\Psi\ra $,
 to connect descriptions $|\Psi\ra $ and $\Pm|\Psi\ra $ in the two FRs,
 it seems inevitable to assume a probabilistic relationship between them.
 We assume that the relationship takes a form
 similar to Born's probabilistic interpretation of wave function.
 The most straightforward generalization of Burn's rule is that,
 with the probability $\| \Pm|\Psi\ra \|^2$,
 the total system is described by $\Pm|\Psi\ra $ in the FR of $\R_{\{\mu\}}$.

 Based on the above analysis, we introduce the last postulate as follows.
 \bi
 \item \textbf{Postulate IV}:
 Suppose that a system $\R$ may have a reference property related to
 $\{\Pm\}$ for a description of the total system given by $\Pm |\ww\Psi\ra $
 with certain value $\mu$ in the FR of $\R_{\{\mu\}}$ for some vector $|\ww\Psi\ra $.
 If the state of the total system is described by $|\ww\Psi\ra $ in a FR of $\R_{\{\nu\}}$,
 then, with a probability $p_\mu$ satisfying $p_\mu \doteq \la \Psi | \Pm | \Psi\ra $,
 the FR of $\R_{\{\mu\}}$ is valid and the total system is described by $\Pm | \Psi\ra $
 in this FR,
 meanwhile, with the probability $p_{\ov\mu}=1-p_\mu $, the total system is still
 described in the FR of $\R_{\{\nu\}}$ but by the vector $\ov \PP_\mu | \Psi\ra $.
 Here, $|\Psi\ra =|\ww\Psi\ra / \| |\ww\Psi\ra \|$ and  $\ov\PP_\mu = 1-\Pm$.
 \ei
 According to the definition of reference property given above,
 the condition stated in Postulate IV, i.e.,
 that $\R$ may have a reference property related to $\{\Pm\}$ for the description
 $\Pm|\Psi\ra $ of the total system  can be expressed in the following explicit way,
 \bey  \label{cond-RS}   \hspace{1.cm}
 \la \Psi_\mu(t') |\Pm |\Psi_\mu(t')\ra  \doteq 1 \ \ \text{for} \ t'\in [t,t+\tau_s],
 \eey
 where
 \bey |\Psi_\mu(t')\ra = \frac {U(t',t) \Pm |\ww \Psi\ra}{\| \Pm |\ww \Psi\ra \|}.   \eey

 In the assignment of probability $p_\mu$ in Postulate IV and in Eq.~(\ref{cond-RS}),
 the symbol ``$\doteq$'' means the relationship of exact or approximate equality.
 In this paper, we use $A\doteq B$ to indicate $|A-B|\le \epsilon_x$ (or
 $\| A-B \| \le \epsilon_x$, or the like),
 where $\epsilon_x \ge 0$ is a small quantity, depending on the situation considered,
 with $x$ indicating the specific situation.
 The parameter $\epsilon_x$ for Eq.~(\ref{cond-RS}) will be written as $\epsilon_R$.
 In  Sec.~\ref{sect-PPA}, we'll show that if one takes the choice $\epsilon_R=0$,
 in most cases the trivial FR of $\R_I$ would be the only valid FR.
 For this reason, we usually assume a small but finite value of $\epsilon_R$.
 The implication of a finite $\epsilon_R$ will also be discussed in Sec.~\ref{sect-PPA}.

 The parameter $\epsilon_x$ in  $p_\mu \doteq \la \Psi | \Pm | \Psi\ra $,
 which we denote by $\epsilon_p$, is not independence of $\epsilon_R$.
 In fact, since $\la \Psi_\mu(t') |\Pm + \ov \PP_\mu |\Psi_\mu(t')\ra =1$,
 from Eq.~(\ref{cond-RS}) and the definition of $\epsilon_R$, we have
 \be \la \Psi_\mu(t') | \ov \PP_\mu |\Psi_\mu(t')\ra \le \epsilon_R . \label{Psi-eR} \ee
 The left hand side of (\ref{Psi-eR}) is the amount of leakage of the vector
 from the subspace of $\Pm$ to its orthogonal subspace.
 Therefore, we assume that $\epsilon_p = \epsilon_R$.
 It is easy to see that Eq.~(\ref{cond-RS}) can be equivalently written as
 \be \ov\PP_\mu U(t',t) \Pm |\Psi\ra \doteq 0 \ \ \ \ \text{for} \ t'\in [t,t+\tau_s],
 \label{cond-RS-ov} \ee
 with $\epsilon_x$ changed accordingly.


 \section{FRs and the principle of consistent description}
 \label{sect-BC}

 The basic postulates introduced above imply that FRs in the theory proposed here
 have some properties not possessed in the usual quantum and classical theories.
 In this section, we discuss some new features of FRs and descriptions.

 \subsection{Condition for the validity of a FR}
 \label{sect-FR-trans}

 Postulate IV gives the condition for a previously valid FR to keep valid.
 It can be obtained by identifying $\{\nu \}$ with $\{\mu\}$ in Postulate IV.
 Specifically, if a FR of $\R_{\{\mu\}}$ is valid initially, in which the state of the
 total system $\R+\E$ is described by a normalized vector $|\Psi_{\mu}(t_0)\ra $
 with a definite value $\mu_0$ of a reference property related to $\{\Pm\}$,
 then, as long as the following condition is satisfied,
 \bey  \label{cond-FR}   \hspace{1.cm}
 \la \Psi(t')| \PP_{\mu_0} |\Psi(t')\ra \doteq 1 \ \ \ \text{for} \ t'\in [t_0,t+\tau_s],
 \eey
 the FR of $\R_{\{\mu\}}$ keeps valid within the time interval $[t_0,t]$.
 Here, $|\Psi(t')\ra = U(t',t_0)|\Psi_{\mu_0}(t_0)\ra $.

 With time passing, as a result of Schr\"{o}dinger evolution,
 a previously valid FR may become invalid.
 In fact, as long as the FR of $\R_{\{\mu\}}$ is valid, time evolution of the total system
 in this FR obeys Schr\"{o}dinger equation.
 However, when the total Hamiltonian may induce transition among different subspaces $\HH_\mu$,
 at some later time, $|\Psi(t)\ra $ will have components in subspaces other than $\HH_{\mu_0}$.
 This is in contradiction with Postulate I.
 To avoid confliction between Postulates I and III, the only solution is that
 the FR of $\R_{\{\mu\}}$ becomes invalid at some later time.

 The condition for the FR of $\R_{\{\mu\}}$ becomes invalid is that
 the Eq.~(\ref{cond-FR}) is unsatisfied.
 When a FR of $\R_{\{\mu\}}$ becomes invalid, Schr\"{o}dinger evolution in the FR
 must stop.
 This cease of Schr\"{o}dinger evolution is not due to any
 limitation of Schr\"{o}dinger equation,
 but because of the invalidity of the employed FR.

 \subsection{Relation between descriptions given in different FRs }

 Postulate IV gives the rule for relating descriptions given in different FRs
 of the same RS $\R$.
 Let us first consider the special case that we have a description $|\Psi\ra $
 of the total system
 in a FR of $\R_{\{\nu\}}$ and want to know the description in the FR of $\R_I$.
 Since Eq.~(\ref{cond-RS}) is always satisfied for the trivial division $\{I\}$,
 according to Postulate IV, with unit probability the total system can be described
 by the same vector $|\Psi\ra $ in the FR of $\R_I$.
 This means that we can always transform the description of the total system given
 in a non-trivial FR to the trivial FR without any change of the description.

 More generally, suppose the total system is described by $|\Psi\ra $ in a FR of $\R_{\{\nu\}}$
 and we want to know the possibility of description given in a FR of  $\R_{\{\mu\}}$.
 If the condition (\ref{cond-RS}) is satisfied for all the values of $\mu$,
 we may transform the description of the total system given in the FR of $\R_{\{\nu\}}$
 to that in the FR of $\R_{\{\mu\}}$.
 However, unlike Galilean or Lorentz transformation discussed in the usual quantum
 and classical mechanics,
 this transformation is a probabilistic one:
 $|\Psi\ra \to \Pm|\Psi\ra $ with the probability $p_\mu$, hence, is irreversible.

 Furthermore, in the case that the condition (\ref{cond-RS}) is satisfied only for some of the
 values of $\mu$, we can not completely transform from the FR of $\R_{\{\nu\}}$
 to the FR of $\R_{\{\mu\}}$, since the later FR is not absolutely valid.
 For example, if Eq.~(\ref{cond-RS}) is satisfied only for one value $\mu$,
 then, as stated in Postulate IV, the description $\ov \PP_\mu |\Psi\ra $
 is given in the original FR of $\R_{\{\nu\}}$;
 only the $\Pm|\Psi\ra $ part is allowed in the FR of $\R_{\{\mu\}}$

 To understand the origin of the above discussed inequivalency in
 descriptions of the same state of the total system in different FRs of $\R$,
 one should note that different reference properties of $\R$ are employed for different FRs.
 Since the values of reference properties give measurement records, a difference in FRs
 implies a difference in the measuring ability of the measuring apparatus,
 hence, a difference in the information that is available from measurement.
 Thus, the information available in different FRs may be different, as a consequence,
 descriptions of the same state of the total system in difference FRs may be
 inequivalent.

 For comparison, we note that measuring apparatuses are not included in the
 description of the usual quantum mechanics.
 This implies that one may in principle assume that each measuring apparatus may perform
 all available measurements with the ultimate accuracy.
 Then, descriptions for the same state in different FRs should contain equivalent
 information, otherwise, measurements may reveal conflicting properties for the same state.
 Therefore, transformation between different FRs should be reversible.
 In classical mechanics, ultimate measuring ability is always assumed,
 hence, one has a result similar to that in the usual quantum mechanics.

 \subsection{Physically equivalent descriptions and the principle of
 consistent description}
 \label{sect-pcd}
 \label{sect-phy-equiv}

 In the usual quantum mechanics, $|\Psi\ra $ and $\xi |\Psi\ra $ represent the
 same physical state for a total system, where $\xi$ is a non-zero complex number.
 Here, an important consequence of the postulates introduced above is that
 more descriptions are possible for the same state of the total system $\R+\E$.

 Let us consider a state of the total system described by a normalized vector $|\Psi\ra $
 in the FR of $\R_I$,
 which satisfies the condition (\ref{cond-RS}) for a value $\mu$ of a division $\{\Pm\}$.
 According to Postulate IV, this implies that with the probability
 $p_\mu$ the state of $\R+\E$ is described by $\Pm |\Psi\ra $ in the FR of
 $\R_{\{\mu\}}$, and with the probability $p_{\ov \mu}$ the state of $\R+\E$
 is described by $\ov \PP_\mu |\Psi\ra $ in the FR of $\R_I$.
 As discussed in the previous section, we can transform
 the description $\Pm |\Psi\ra $ given in the FR of $\R_{\{\mu\}}$ to the same vector
 in the FR of $\R_I$.

 Then, the state of the total system has the following density operator
 description in the FR of $\R_I$, with an ordinary ensemble interpretation:
 \be \rho \doteq \Pm |\Psi\ra \la \Psi| \Pm + \ov \PP_\mu |\Psi\ra \la \Psi| \ov \PP_\mu .
 \label{rho-ori} \ee
 If the vector $|\Psi\ra $ satisfies Eq.~(\ref{cond-RS}) for all the values of $\mu$,
 the total system can be described by
 \be \rho \doteq \sum_\mu \Pm |\Psi\ra \la \Psi| \Pm . \label{rho-ori2} \ee
 Thus, in the FR of $\R_I$, besides $|\Psi\ra $, the total system also has a density operator
 description $\rho$.

 The two description $|\Psi\ra $ and $\rho$ must be physically equivalent,
 otherwise, the theory would be inconsistent.
 Here, two descriptions for the same state given in the same FR
 are said to be \emph{physically equivalent},
 if they give the same predictions for results of measurements on the system.
 In the theory here, predictions for measurement results mean probabilities for
 $\R$ to take  possible values of its reference properties.
 It is easy to see that if two descriptions $A$ and $B$ are physically equivalent,
 meanwhile descriptions $B$ and $C$ are also physically equivalent, then,
 the two descriptions $A$ and $C$ must give the same predictions for measurement results
 and be physically equivalent, too.
 Therefore, physically equivalent descriptions form an equivalent class.

 We express the above consistency requirement for the theory as a principle.
 \bi
 \item  \emph{The principle of consistent description}:
 Descriptions given in the same FR for the same state of a system
 must be physically equivalent.
 \ei
 This principle is a requirement that should be satisfied in every physical theory.
 It is usually not emphasized, because in most physical (quantum and classical) theories
 it can be easily proved and does not lead to any important consequence.
 However, the physical equivalence of $|\Psi\ra $ and $\rho$ is clearly non-trivial,
 hence, it is reasonable to expect that this principle play a non-trivial role here.

 The physical equivalence between $|\Psi\ra $ and $\rho$
 is one of the major differences of the theory proposed here from
 the usual quantum mechanics.
 In the usual quantum mechanics, a density operator of the form in Eq.~(\ref{rho-ori})
 is called a mixed state, while a vector $|\Psi\ra $ is called a pure state;
 there, the two types of description can not be physically equivalent, since their
 difference can in principle be tested experimentally by an outside measuring apparatus.
 The possibility for a pure vector description to be physically equivalent
 to a density operator description in the theory here lies in the fact that
 the measuring apparatus (RS here) is a part of the described total system.
 This feature implies more limitation to experimentally obtainable information
 in the theory proposed here than in the usual quantum mechanics.
 This limitation is in fact related to the triple roles played by RS
 discussed in Introduction,
 namely, of a ``describer'', a participant, and a ``judge'' \cite{note-bohr}.

 Finally, we give a discussion for methods of constructing physically equivalent descriptions
 from a given normalized vector $|\Psi\ra$.
 For this purpose, we need to consider the FR of $\R_I$ only, since
 descriptions given in a non-trivial FR of $\R$ can be transformed without
 any change to the FR of $\R_I$.

 First, from Postulate IV, it is easy to see that, as in the usual quantum mechanics,
 $|\Psi\ra $ for the total system is physically equivalent to $\xi |\Psi\ra $.
 Second, one may construct $\rho$ in Eq.~(\ref{rho-ori}) from $|\Psi\ra $.
 Finally, if the division $\{\Pm\}$ discussed above is a non-trivial one, then,
 $|\Psi\ra $ is physically equivalent to the following vector
 \be |\Psi'\ra = e^{i\theta_1} \Pm |\Psi\ra  + e^{i\theta_2} \ov \PP_\mu |\Psi\ra , \ee
 where $\theta_1$ and $\theta_2$ are arbitrary real numbers.
 In fact, if $|\Psi\ra $ satisfies Eq.~(\ref{cond-RS}) for a value $\mu$ of $\{\mu\}$,
 then, $|\Psi '\ra $ satisfies Eq.~(\ref{cond-RS}) for the same value $\mu$.
 Therefore, $|\Psi'\ra $ is physically equivalent to $\rho$ in Eq.~(\ref{rho-ori}),
 hence,  physically equivalent to $|\Psi\ra $.
 In these three methods, the first one is the same as that in the usual quantum mechanics,
 and the third one is in fact an application of the second one.
 Therefore, basically, the second method is the new one that needs further discussion.

 \section{Implication of the principle of consistent description
 in physically allowed vectors}
  \label{sect-CC}

 In this section, we show that the principle of equivalent description imposes restriction
 to vectors in the Hilbert space that can be associated with physical states.
 For this, we'll first discuss time evolution described in the FR of $\R_I$, then,
 derive an expression for the principle.

 \subsection{Descriptions given in the FR of $\R_I$}
 \label{sect-FR-I}

 A pure vector description of the total system has the simplest time evolution, namely,
 Schr\"{o}dinger evolution.
 However, as discussed above, a pure vector description may be physically equivalent
 to a density operator description.
 When this possibility is taken into account, time evolution of the total system
 in the FR of $\R_I$ becomes much more complex and we discuss it in this section.

 Let us start from an initial state of $\R+\E$, which is described by
 a normalized vector $|\Psi(t_0)\ra$.
 (Generalization of the following discussions to the case of an initial density
 operator is straightforward.)
 Suppose $|\Psi(t_0)\ra$ satisfies the condition
 (\ref{cond-RS}) for a value $\mu^{(0)}$ of a division $\{\PP_{\mu^{(0)}}\}$
 with $|\ww\Psi\ra $ taken as $|\Psi(t_0)\ra$.
 For the same reason as writing $\rho$ in Eq.~(\ref{rho-ori}),
 the initial state can be described equivalently by the following density operator,
 \be \rho(t_0) \doteq \sum_{m^{(0)}} |\Psi_{m^{(0)}}(t_0)\ra \la\Psi_{m^{(0)}}(t_0)|,
 \label{r-t0} \ee
 where $m^{(0)}=1^{(0)}, 2^{(0)}$ and $|\Psi_{m^{(0)}}(t_0)\ra = \PP_{m^{(0)}} |\Psi(t_0)\ra$,
 with
 \bey \left . \PP_{m^{(0)}}\right |_{m=1} = \PP_{\mu^{(0)}}, \ \
 \left . \PP_{m^{(0)}}\right |_{m=2} = \ov\PP_{\mu^{(0)}} . \label{Pm0}
 \eey
 That is, the total system is described by
 $|\Psi_{m^{(0)}}(t_0)\ra$ with the probability $\la\Psi_{m^{(0)}}(t_0) |\Psi_{m^{(0)}}(t_0)\ra$.
 Clearly, $ |\Psi(t_0)\ra =  \sum_{m^{(0)}} |\Psi_{m^{(0)}}(t_0)\ra$.

 \begin{figure}[!t]
  \includegraphics[width=\columnwidth]{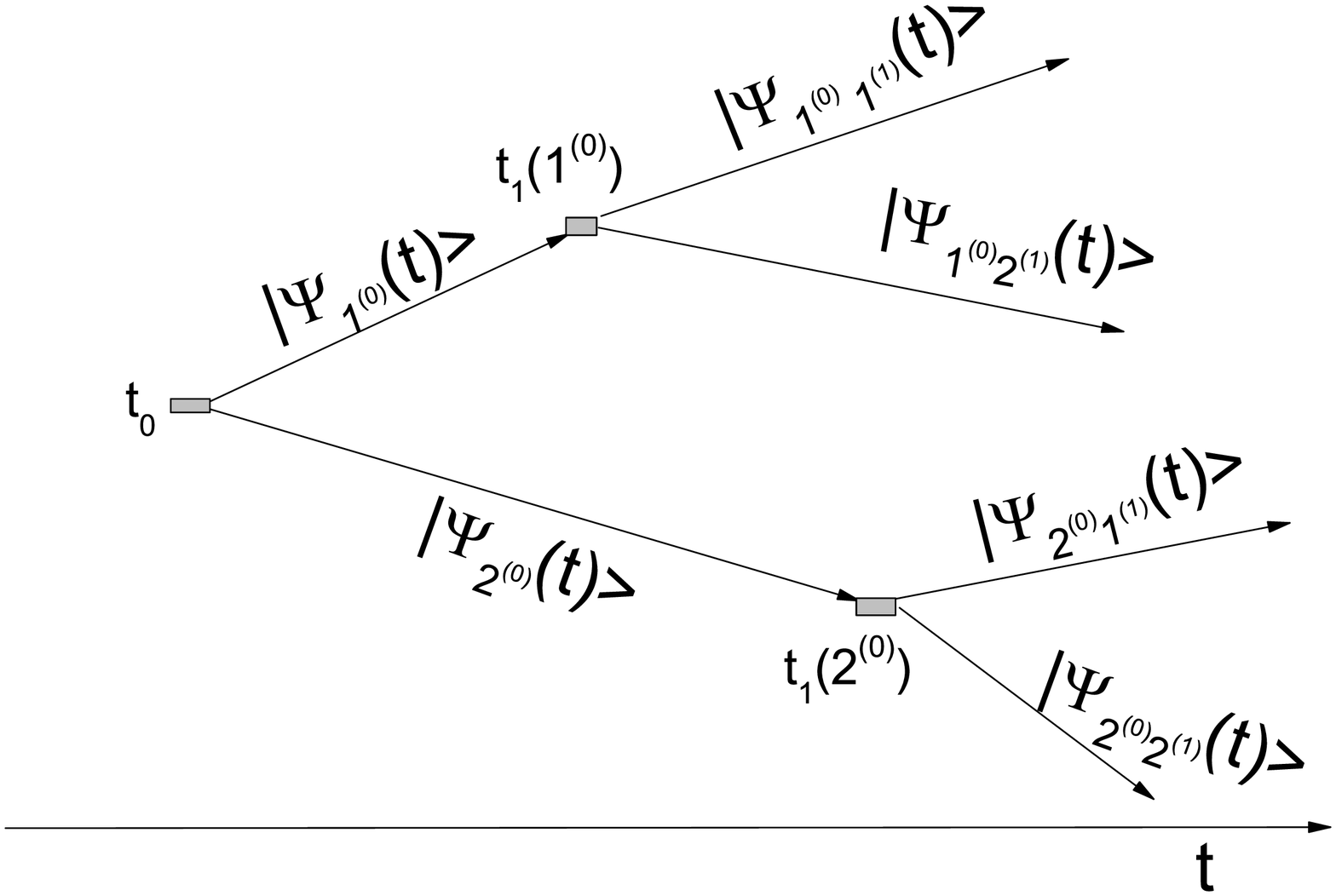}
   \vspace{-0.4cm}
 \caption{Schematic plot for the tree structure of the time evolution of the components of
 a density operator description $\rho(t)$ of the total system $\R+\E$.
 It starts from one initial vector $|\Psi(t_0)\ra $, which is indicated by a small square labeled
 by $t_0$, and splits into two components $|\Psi_{m^{(0)}}(t)\ra $ of $m=1,2$.
 The next two small squares indicate the splitting points of the two components
 $|\Psi_{m^{(0)}}(t)\ra $ at the times $t_1(m^{(0)})$ of $m=1$ and 2, respectively.
 A component $|\Psi_{m^{(0)}}(t)\ra $ between two successive times $t_0$ and $t_1(m^{(0)})$
 is represented by a short line with time direction and is called a branch.
 At the two times $t_1(m^{(0)})$ of $m=1$ and 2,
 the two components $|\Psi_{m^{(0)}}(t_1)\ra $ split again,
 each into two sub-branches $|\Psi_{m^{(0)}m^{(1)} }(t)\ra $.
 We call a sequence of splitting, $(m^{(0)},m^{(1)}\ldots )$, a path of splitting.
 } \label{fig-bra}
 \end{figure}

 From the description $|\Psi(t_0)\ra $ to its equivalent description $\rho(t_0)$,
 one may say that the vector $|\Psi(t_0)\ra$ ``splits'' into
 two components $|\Psi_{m^{(0)}}(t_0)\ra$.
 Each component $|\Psi_{m^{(0)}}(t_0)\ra$ evolves independently
 in the FR of $\R_I$, obeying Schr\"{o}dinger equation,
 \be |\Psi_{m^{(0)}}(t)\ra  = U(t,t_0)|\Psi_{m^{(0)}}(t_0)\ra . \label{Psi-m0} \ee
 These features can be plotted schematically as shown in the left part of Fig.~\ref{fig-bra}.

 If a component $|\Psi_{m^{(0)}}(t)\ra$ in Eq.~(\ref{Psi-m0})
 satisfies Eq.~(\ref{cond-RS}) for a value
 $\mu^{(1)}$ of a division $\{\PP_{\mu^{(1)}}\}$ at a time $t_1$,
 then, we can do the same thing for $|\Psi_{m^{(0)}}(t_1)\ra$ as done above for $|\Psi(t_0)\ra $.
 Here, $t_1$ is a function of $m^{(0)}$, $t_1=t_1(m^{(0)})$.
 As a result, the component $|\Psi_{m^{(0)}}(t_1)\ra $ is
 effectively equivalent to the following density operator
 \be \sum_{m^{(1)}} |\Psi_{m^{(0)} m^{(1)}}(t_1)\ra \la\Psi_{m^{(0)} m^{(1)}}(t_1)| ,
 \label{sum-m01} \ee
 where
 \bey  |\Psi_{m^{(0)} m^{(1)}}(t_1)\ra =   \PP_{m^{(1)}}|\Psi_{m^{(0)}}(t_1)\ra .
 \label{Psi-m01} \eey
 Here, $ \PP_{m^{(1)}}$ is defined in a way similar to $ \PP_{m^{(0)}}$ in Eq.~(\ref{Pm0}).
 Thus,  the component $|\Psi_{m^{(0)}}(t_1)\ra $ splits into
 sub-components $|\Psi_{m^{(0)} m^{(1)}}(t_1)\ra$.
 Beyond $t_1$, each  component $|\Psi_{m^{(0)} m^{(1)}}(t_1)\ra$ has Schr\"{o}dinger
 evolution again.
 These features are also plotted in Fig.~\ref{fig-bra}.
 A component $|\Psi_{m^{(0)}}(t)\ra$ of $t\in(t_0,t_1)$ is represented by a short line,
 which we call a \emph{branch}.

 With increasing time, the splitting of components (branches) may happen again and again.
 We use $\alpha_n$ to indicate a sequence of $n+1$ splitting and call it
 a \emph{path} of $n+1$ steps of splitting,
 \be \alpha_n \equiv \left ( m^{(0)}(t_0) \to m^{(1)}(t_1) \to
 \ldots \to m^{(n)}(t_n) \right ).
 \label{path-mu} \ee
 The splitting in $\alpha_n$ happen at the times $t_0,t_1,\ldots ,t_n$
 and generate new components indicated by $m^{(0)}, m^{(1)}, \ldots, m^{(n)}$, respectively.
 A splitting time $t_n$ is a function of the previous path, therefore, it can be
 written as $t_n(\alpha_{n-1})$.
 Sometimes, it may be more convenient not to specify the number of the steps of a path and,
 in this case, we indicate a path simply by $\alpha$.

 The branches and splitting points indicated by
 short lines and small squares in Fig.~\ref{fig-bra} form a structure like a tree.
 For this reason, we call one set of compatible splitting points and branches a \emph{tree}
 and denote it by $\Upsilon$.
 Using expressions like Eqs.~(\ref{Psi-m0}) and (\ref{Psi-m01}),
 it is easy to get the following explicit expression for the component
 obtained through a path $\alpha_n \in \Upsilon$,
 \bey \nonumber |\Psi_{\alpha_n}(t)\ra = U(t,t_n) \PP_{m^{(n)}} U(t_{n},t_{n-1}) \PP_{m^{(n-1)}}
 \\ \cdots  \PP_{m^{(1)}} U(t_1,t_0) \PP_{m^{(0)}} |\Psi(t_0)\ra ,
 \label{Psi-alpha} \eey
 where $\PP_{m^{(i)}}$ are defined in a way similar to $\PP_{m^{(0)}}$ in Eq.~(\ref{Pm0}),
 \bey \left . \PP_{m^{(i)}}\right |_{m=1} = \PP_{\mu^{(i)}}, \ \
 \left . \PP_{m^{(i)}}\right |_{m=2} = \ov\PP_{\mu^{(i)}} . \label{Pmi}
 \eey
 Then, the state of the total system at the time $t$ can be written in the following
 density-operator form in the FR of $\R_I$,
 \be \rho_\Upsilon(t) 
 = \sum_{\alpha \in \Upsilon} |\Psi_{\alpha}(t)\ra \la \Psi_{\alpha}(t)|.
 \label{rho-fi} \ee
 The probability for the realization of a path $\alpha$ is
 \be P_{\alpha}(t) = \la \Psi_{\alpha}(t) |\Psi_{\alpha}(t)\ra . \label{Pan} \ee
 Using the relation $ \sum_{m^{(i)}} \PP_{m^{(i)}} =1 \label{Pmn=1} $,
 it is easy to verify that
 \be |\Psi(t)\ra = \sum_{\alpha \in \Upsilon } |\Psi_\alpha (t)\ra .  \label{PUP1} \ee

 It is straightforward to generalize the above results to the generic case of splitting.
 For example, let us consider the splitting of
 a component $|\Psi_{\alpha}(t)\ra$ with $t>t_{n-1}$, which is reached through a path
 $\alpha$ with $n$ steps of splitting.
 At a time $t=t_n$, the component may satisfy the condition (\ref{cond-RS})
 for $(L-1)$ values of $\mu^{(n)}$ for a division $\{\PP_{\mu^{(n)}}\}$.
 We use $s^{(n)}$ to denote the set of these $(L-1)$ values of $\mu^{(n)}$.
 In this generic case, we define the projection operators $\PP_{m^{(n)}}$ as follows:
 \bey \PP_{m^{(n)}} = \left \{
 \begin{array}{ll}
 \PP_{\mu^{(n)}\in s^{(n)}},  \ \ m=1,\ldots , L-1,
 \\ \ov \PP_{s^{(n)}}, \hspace{1.0cm} \ m=L ,
 \end{array}
 \right . \eey
 where, for $\PP_{m^{(n)}}$ with different values of $m$ smaller than $L$,
 we choose different $ \mu^{(n)} \in s^{(n)}$, and
 \be \ov \PP_{s^{(n)}} = 1 - \sum_{\mu^{(n)} \in s^{(n)}} \PP_{\mu^{(n)}}. \ee
 It is easy to see that in this generic case Eqs.~(\ref{Psi-alpha}) and
 (\ref{rho-fi})-(\ref{PUP1}) remain valid, and more than two short lines may come
 out of each small square in Fig.~\ref{fig-bra}.

 It is important to note that
 the above discussed process of splitting is not unique.
 In fact, since a splitting may happen whenever Eq.~(\ref{cond-RS}) is satisfied,
 there are usually many ways of choosing the splitting, resulting in many different
 trees $\Upsilon$.
 (For this reason, $\alpha_{n-1}$ does not completely fix $t_n$.)
 Thus, $|\Psi(t)\ra $ is physically equivalent to many different density operators.

 For a fixed time, $\rho_\Upsilon(t)$ {may supply more information than} $|\Psi(t)\ra $.
 Indeed, if a component $|\Psi_\alpha(t)\ra $ of $\alpha \in \Upsilon$
 satisfies Eq.~(\ref{cond-RS}) for a value $\mu$ of a division $\{\Pm\}$
 while another component $|\Psi_{\alpha'}(t)\ra $ of $\alpha' \in \Upsilon$ does not,
 then, the total vector $|\Psi(t)\ra $ does not satisfy Eq.~(\ref{cond-RS})
 for this value $\mu$ of $\{\Pm\}$.
 In this case, the property $\{\mu\}$ may be a reference property for
 the branch $|\Psi_\alpha(t)\ra $, while it can not be a reference property
 for the total vector $|\Psi(t)\ra $.
 However, when the time evolution is considered,
 $\rho_\Upsilon(t)$ does not supply more information than $|\Psi(t)\ra $.
 This is because, due to the uniqueness of the solution of Schr\"{o}dinger equation,
 the initial vector $|\Psi (t_0)\ra $ can be determined
 from $|\Psi(t)\ra $, hence, it is in principle possible
 to get $\rho_{\Upsilon}(t)$ from $|\Psi(t)\ra $.


 \subsection{Consistency of predictions for measurement results}
 \label{sect-cdes}

 In this section, we first show that a vector description $|\Psi\ra $
 gives consistent predictions for measurement results related to different possible
 reference properties.
 Then, we discuss relation between predictions given by different types of
 descriptions, pure vector and density operator, and
 derive an explicit expression of the principle of consistent description.
 It would be sufficient to discuss descriptions given in the trivial FR of $\R_I$.
 In fact, as discussed in Sec.~\ref{sect-FR-trans}, each description given in a non-trivial
 FR of $\R$ can be transformed to the FR of $\R_I$ without any change.

 \subsubsection{Predictions given by the same vector description}

 We discuss here predictions given by the same pure-vector description
 of the total system for measurement results taking the form of different reference properties.
 Let us consider a normalized vector description $|\Psi\ra $
 of the total system at a time $t$ in the FR of $\R_I$.

 Suppose there may exist two reference properties for this vector description,
 say, related to $\{\PP_{\mu}\}$ and $\{\PP_{\nu}\}$.
 Specifically, $|\Psi\ra $ satisfies Eq.~(\ref{cond-RS}) for a value $\mu$ of $\{\Pm\}$,
 hence, according to Postulate IV, there is a probability $p_\mu \doteq \| \Pm|\Psi\ra \|^2$
 for $\R+\E$ to be described by a vector $\PP_\mu |\Psi\ra$ in the FR of $\R_{\{\mu\}}$.
 Similarly, $|\Psi\ra $ also satisfies Eq.~(\ref{cond-RS}) for a value $\nu$ of $\{\Pnn\}$,
 which can be equivalently written as [see Eq.~(\ref{cond-RS-ov})]
 \be \ov \PP_\nu U(t',t) \Pnn |\Psi\ra \doteq 0 \ \ \text{for} \
 t'\in [t,t+\tau_s],  \label{cond-RS-nu} \ee
 and there is a probability $p_\nu \doteq \| \Pnn |\Psi\ra \|^2 $ for
 $\R+\E$ to be described by $\PP_\nu |\Psi\ra$ in the FR of $\R_{\{\nu\}}$.

 Suppose the subspace $\HH_\nu $ is included in $\HH_\mu$, $\HH_\nu \subset \HH_\mu$.
 (The case of $\HH_\nu \supset \HH_\mu$ can be treated similarly.
 In the case that neither of the two subspaces is included in the other,
 there is no definite relation between $p_\nu$ and $p_\mu$.)
 In this case,  $\Pnn \subset \Pm$, as a result, $\Pnn \Pm =\Pnn$.
 Using this result, Eq.~(\ref{cond-RS-nu}) can be written as
 \be  \ov \PP_\nu U(t',t) \Pnn (\Pm |\Psi\ra ) \doteq 0 \ \ \text{for} \
 t'\in [t,t+\tau_s].
 \label{Pmn-H-0} \ee
 This implies that when the total system is described in the FR of $\R_{\{\mu\}}$
 by $\PP_\mu |\Psi\ra$, according to Postulate IV, the total system has
 a reference property $\nu$ with the probability
 \be p_{\mu \nu} = \frac {\la \Psi |\Pm \Pnn \Pm |\Psi\ra }{\| \PP_{\mu} |\Psi\ra \|^2}. \ \
 \ee
 Then, using $\Pnn \subset \Pm$ again, we see that $p_\nu = p_{\mu \nu} p_\mu$.
 Therefore, the predictions of $|\Psi\ra $ for measurement results recorded as the values of
 the two properties $\{\mu\}$ and $\{\nu\}$  are consistent.

 \subsubsection{An expression of the principle of consistent description}
 \label{sect-epcd}

 Now, we discuss the relation between predictions of physically
 equivalent descriptions for the same state of the total system in the FR of $\R_I$.
 As discussed in Sec.~\ref{sect-phy-equiv}, there exist three methods of constructing
 physically equivalent descriptions for a given pure vector description.
 The first one, from $|\Psi\ra $ to $\xi |\Psi\ra $, is a trivial one;
 and the third one is in fact an application of the second one.
 Therefore, here we need to consider only the relation established by the second method,
 i.e., the relation between $|\Psi(t)\ra $ and $\rho_\Upsilon(t)$ discussed
 in Sec.~\ref{sect-FR-I}.



 Suppose $|\Psi(t)\ra$ satisfies Eq.~(\ref{cond-RS}) for a value $\mu$ of the division $\{\Pm\}$.
 According to Postulate IV, the probability for $\R$ to have the value $\mu$ of the corresponding
 reference property, denoted by $P_\Psi(\mu,t)$, can be written in the following form,
 \be P_\Psi (\mu,t) \doteq  {\sum_{\alpha \alpha'}}  \la \Psi_{\alpha'}(t)|\Pm | \Psi_{\alpha}(t)\ra ,
 \label{wP} \ee
 where Eq.~(\ref{PUP1}) has been used.
 Due to the finiteness of $\tau_s$,
 when $|\Psi(t)\ra$ satisfies Eq.~(\ref{cond-RS}) for the value $\mu$,
 all the components $|\Psi_\alpha(t)\ra $ in Eq.~(\ref{PUP1}) should
 satisfy Eq.~(\ref{cond-RS}) for the same value $\mu$ of the division $\{\Pm\}$.
 Then, $\rho_\Upsilon (t)$ in Eq.~(\ref{rho-fi}) predicts the following
 probability for $\R$ to have the definite value $\mu$ of the
 corresponding reference property $\{\mu\}$,
 \be P_\Upsilon(\mu,t) \doteq  {\sum_{\alpha}}  \la \Psi_{\alpha}(t)|\Pm | \Psi_{\alpha}(t)\ra .
 \label{P-Up} \ee

 The principle of consistent description  requires that
 \be P_\Psi(\mu,t) \doteq P_\Upsilon(\mu,t) . \label{wPPU} \ee
 Equations (\ref{wPPU}), (\ref{wP}), and (\ref{P-Up}) imply that
 \bey \label{D-diag} \D_{\alpha \alpha'}^\mu \doteq  \delta_{\alpha \alpha'}
 \D_{\alpha \alpha}^\mu
 \ \ \ \forall \alpha , \alpha' \in \Upsilon, \eey
 where
 \bey  \D_{\alpha \alpha'}^\mu \equiv \la \Psi_{\alpha}(t)|\Pm | \Psi_{\alpha'}(t)\ra .
 \label{Da} \eey
 Note that $\sum_\mu \D_{\alpha \alpha}^\mu = P_\alpha$ is
 the probability for the realization of a path $\alpha$.
 Equation (\ref{D-diag}) may hold for $\alpha \ne \alpha'$ due to decoherence effects.
 We'll give further discussion for this point in a simple model
 of RS in Sec.~\ref{sect-model}.

 It is not difficult to show that physically equivalent descriptions constructed
 by the other two methods do not lead to new requirements in the FR of $\R_I$.
 Moreover,
 descriptions given in FRs of $\R$ other than the FR of $\R_I$ can be treated in a
 similar way and, in these cases, the principle of consistent description requires
 some relations that are special cases of the relations derived above in the FR of
 $\R_I$.

 To summarize, we reach the following explicit expression for the principle of
 consistent description:
 \bi
 \item For each two descriptions $|\Psi(t)\ra $ and $\rho_\Upsilon(t)$ of the total system
 given in the FR of $\R_I$,
 which are obtained from the same initial condition $|\Psi(t_0)\ra $,
 if $|\Psi(t)\ra $ satisfies Eq.~(\ref{cond-RS}) for a value $\mu$ of a division
 $\{\Pm\}$, then, Eq.~(\ref{D-diag}) must hold.
 \ei

 \subsection{The parameter $\epsilon_R$}
 \label{sect-PPA}
 \label{sect-comple}

 An important feature of the theory proposed in this paper is that the parameter
 $\epsilon_R$ in Eq.~(\ref{cond-RS}) usually takes a small but finite value.
 We discuss the necessity of choosing a finite $\epsilon_R$ in this section, not in
 Sec.~\ref{sect-Postulate-IV} where it is introduced, because $\epsilon_R$ is also
 of relevance to Eq.~(\ref{D-diag}), an expression of the principle of consistent description.

 Let us first consider the ideal case of $\epsilon_R=0$, which implies $\epsilon_p=0$
 with zero error in the assignment of probability.
 In this ideal case, all the symbol $\doteq$ appearing in the equations given above can be
 replaced by the exact equality.
 However, in this ideal case, one faces the following problem:
 The interaction Hamiltonian $H_I$ usually generates non-zero coupling among
 different subspaces $\HH_{\mu}$ for each non-trivial division $\{\Pm\}$.
 Then, the time evolution of an arbitrary initial vector can satisfy Eq.~(\ref{cond-RS})
 for a non-trivial division only quite occasionally.
 This implies that the trivial division $\{I\}$ would usually be the only one,
 for which Eq.~(\ref{cond-RS}) can be satisfied.
 (Here, for the simplicity in discussion,
 we do not discuss the case that the total interaction Hamiltonian  has some symmetry
 is commutable with projection operators of some non-trivial division.
 It is in principle possible to extend the discussions given here
 to the case of a $H$ with additional symmetry.)

 One faces a similar problem with Eq.~(\ref{D-diag}).
 In fact, the only known mechanism for Eq.~(\ref{D-diag}) to hold seems decoherence
 induced by environment.
 As well known, decoherence may make an off-diagonal element extremely small, but
 can hardly make it zero.
 Therefore, for Eq.~(\ref{D-diag}) to hold with exact equality,
 usually the trivial division $\{I\}$ is the only choice.

 It is not difficult to see that, compared with the choice of $\epsilon_R=0$,
 some non-trivial reference properties may become possible
 if we take a small but finite value of $\epsilon_R$.
 Therefore, in order to explain the fact that we can perform measurements
 in the many situations done in laboratories, which requires non-trivial reference
 properties of RS, one should choose non-zero $\epsilon_R$.
 A small but finite $\epsilon_R$ implies a small but finite error in the assignment of
 the probability $p_\mu $ in Postulate IV.
 This type of error is inherent in the theory with a finite $\epsilon_R$,
 hence, we may call it an \emph{intrinsic error}.
 We remark that the intrinsic error is controllable and can be made as small as one likes.

 Thus, the proposed theory has the following unusual feature:
 On one hand, with zero $\epsilon_R$,
 the theory has zero intrinsic error, but, it has a very weak prediction ability
 for measurement results.
 On the other hand, with finite $\epsilon_R$, the theory may be more effective in predicting
 measurement results, but, at the cost of non-zero intrinsic error.
 It is impossible to achieve the ultimate accuracy with zero intrinsic error,
 meanwhile, have a good ability of predicting measurement results.
 This feature of the theory has its origin in the triple roles played by RS,
 which have been mentioned in Introduction, and is in spirit somewhat similar to
 N.~Bohr's idea expressed his principle of complementarity.

 \subsection{Allowed $\HH$-region}
 \label{sect-allowed-H}

 The most significant consequence of the principle of consistent description is that
 it imposes a restriction to vectors in the Hilbert space.
 That is, usually not all vectors in $\HH$ can satisfy
 Eq.~(\ref{D-diag}) when they are used as initial vectors.

 Let us consider a set of divisions of $\HH_\R$,
 which we regard as being possibly related to reference properties of $\R$, denoted by  $W_d$,
 \be \label{Wd} W_d =\left \{ \{\Pm \}, \{ \Pnn \}, \ldots \right \}.  \ee
 For each $W_d$, the principle of consistent description selects a special region in the
 Hilbert space, which we call allowed $\HH$-region in what follows.
 \bi \item
 The \emph{allowed $\HH$-region of $W_d$} is composed of those vectors $|\Psi\ra \in \HH$,
 for which Eq.~(\ref{D-diag}) is satisfied for all trees $\Upsilon$ given by divisions
 in $W_d$ when $\frac 1{\| |\Psi\ra\|^{-1}}|\Psi\ra $ are taken as initial vectors.
 \ei
 Clearly, \emph{vectors not in the allowed $\HH$-region of $W_d$ can not be
 associated with any physical state}, when $W_d$ is the set of divisions that are
 possible to be related to reference properties.
 For brevity, we say that vectors outside the allowed $\HH$-region are not allowed physically.

 The simplest $W_d$ is the trivial division $\{I\}$, $W_d = \{I\}$.
 For this $W_d$, since $\ov \PP_I =0$, Eq.~(\ref{D-diag})
 is satisfied for all vectors in the Hilbert space $\HH$.
 Therefore, the allowed $\HH$-region of $W_d=\{I\}$ is just the total Hilbert space.
 Furthermore, it is easy to verify that, for a set $W_d$ not including
 the trivial division $\{I\}$, the set $W_d'=\{ W_d, \{I\} \}$ has the same
 allowed $\HH$-region as $W_d$.

 For a set $W_d$ including a non-trivial division,
 the allowed $\HH$-region is usually smaller than the total Hilbert space.
 To illustrate this, we may consider a simple case that $W_d$ is composed of
 one non-trivial division $\{ \Pm \}$.
 Let us consider an initial vector $|\Psi(t_0)\ra $ in the FR of $\R_I$,
 which lies in a subspace $\HH_{\mu}$ with $\mu=\mu_0$.
 For some later time $t$,
 Schr\"{o}dinger evolution of the vector, $|\Psi(t)\ra =U(t,t_0)|\Psi(t_0)\ra$,
 may spread over more than one subspaces $\HH_\mu$.
 Suppose $|\Psi(t)\ra$ has a time reversal vector in view of Schr\"{o}dinger equation,
 denoted by $|\Psi^{\rm re}(t_0')\ra $ with a relabeling of the time.
 Usually, $|\Psi^{\rm re}(t_0')\ra $ also spreads over more than one subspaces $\HH_\mu$.

 Now, we take $|\Psi^{\rm re}(t_0')\ra $ as a new initial vector.
 Since $|\Psi^{\rm re}(t_0')\ra $ is a time reversal of the vector $|\Psi(t)\ra$,
 at least in some cases, $|\Psi^{\rm re}(t')\ra =U(t',t_0')|\Psi^{\rm re}(t_0')\ra \ra$
 may return back to the subspace $\HH_{\mu_0}$, where $t'=t_0'+t-t_0$.
 Suppose $|\Psi^{\rm re}(t_0')\ra $ satisfies Eq.~(\ref{cond-RS})
 for some value $\mu$ of the division $\{\PP_\mu \}$ and
 $|\Psi^{\rm re}(t')\ra $ satisfies Eq.~(\ref{cond-RS}) for $\mu=\mu_0$.
 Then, from $|\Psi^{\rm re}(t_0')\ra $ to a later density operator description
 $\rho^{\rm re}(t')$, there exist more than one paths from $t_0'$ to $t'$.
 Coherence among these paths is necessary for the vector $|\Psi^{\rm re}(t_0')\ra $,
 which spreads over more than one subspaces $\HH_\mu$,
 to unitarily evolve into $|\Psi^{\rm re}(t')\ra$ in one subspace $\HH_{\mu_0}$.
 This implies that  Eq.~(\ref{D-diag}) can not hold for the time $t'$ when
 $|\Psi^{\rm re}(t_0')\ra $ is taken as the initial vector,
 since this equation means decoherence due to the difference in paths.
 As a result, such a vector $|\Psi^{\rm re}(t_0')\ra $ does not lie in the allowed $\HH$-region
 of $W_d =\{ \Pm \}$.
 Therefore, the allowed $\HH$-region of this $W_d$ is smaller than the total Hilbert space $\HH$.

 \subsection{Reference property and allowed $\HH$-region}

 It is in principle possible to consider any set $W_d$ of divisions of $\HH_\R$
 when discussing reference properties of $\R$.
 However, not each set of $W_d$ is of practical interest, since some of them do not
 have an allowed $\HH$-region that is sufficiently large for explaining the experimental
 results.
 In the extreme case, some may have an empty allowed $\HH$-region.
 Therefore, in describing the total system, it is important to choose a set $W_d$
 possessing a sufficiently large allowed $\HH$-region.

 Different sets $W_d$ may have different allowed $\HH$-regions, therefore, when describing
 the total system, it is also important to fix the employed set of $W_d$.
 Since the set $W_d$ fixes the set of recordable properties of the measuring apparatus
 that are to be considered,
 a difference in the set of $W_d$ implies the possibility of a difference
 in measurement results that can be recorded, hence, a difference in the
 description of the total system.

 Descriptions given under compatible sets of $W_d$ may be related.
 For example, if two sets $W_d$ and $W_d'$ share a common division $\{\Pm\}$ and
 a vector $|\Psi_\mu\ra $ belongs to the allowed $\HH$-regions of both sets,
 then, when the total system is described by $|\Psi_\mu\ra$ in the FR of $\R_{\{\mu\}}$
 under the set $W_d$, it can be described by the same vector in the FR
 of $\R_{\{\mu\}}$ under the set $W_d'$.

 The stability of reference property stated in Postulate I imposes a restriction
 to divisions $\{\Pm\}$ that are of practical interest.
 For example, as an ideal case, we may consider the case that $H_\R$ does not induce
 transition among the subspaces $\HH_\mu$, i.e.,
 \be  \Pm H_{\R} \Pmp = \left ( \Pm H_{\R} \Pm \right ) \delta_{\mu \mu' },
 \ \forall  \mu,\mu' . \label{sta} \ee
 In this case, a reference property is absolutely stable when the interaction Hamiltonian
 $H_I$ can be neglected.
 For brevity, we may call Eq.~(\ref{sta}) a \emph{stability condition} for division.
 (In a practical case, the stability condition may hold approximately.)
 It is easy to verify that Eq.~(\ref{sta}) is equivalent to
 $ [\Pm , H_{\R}] =  0 $ for all $\mu $.

 \section{Applications }
 \label{sect-app}

 In this section, we discuss some of the applications of the theory proposed above.
 We first discuss descriptions given in non-trivial FRs of $\R$.
 Then, making use of the results obtained, we discuss measurement processes and derive
 the axiom of measurement in the standard formalism of quantum mechanics.
 Finally, we show that time evolution of the total system is generally irreversible.

 \subsection{Description given in non-trivial FRs }
 \label{sect-fix-FR}

 Sometimes, one may be interested in description of the total system given in some
 non-trivial FR(s) of $\R$, but not in the trivial FR of $\R_I$.
 In this case, the description is usually given for some sequence of
 non-overlapping time intervals, because
 there may exist some time intervals within which no FR of interest is valid.
 We use $[t_i,t_i^e]$ to denote a time interval within which a FR of
 $\R_{\{\mu^{(i)}\}}$ is valid,
 with $i=0,1, \ldots $ and ``e'' standing for ``end''.
 The initial state is described by $|\Psi_{\mu^{(0)}}(t_0)\ra$,

 Within a time interval $[t_i,t_i^e]$, $\R$ has a definite value $\mu^{(i)}$ of the reference
 property $\{\mu^{(i)}\}$  in the FR of $\R_{\{\mu^{(i)}\}}$.
 For the simplicity in discussion, we write the state of the total system within this
 time interval as a vector, $|\Psi_{\mu^{(i)}}(t)\ra$, which obeys Schr\"{o}dinger equation,
 \be |\Psi_{\mu^{(i)}}(t)\ra = U(t,t_i)|\Psi_{\mu^{(i)}}(t_i)\ra, \qquad t\in [t_i,t_i^e] . \ee
 Within a time interval $(t_i^e, t_{i+1})$, no valid FR of $\R$ is of interest,
 hence, there is a jump for the description of the total system from $t_i^e$
 to $t_{i+1}$.

 To find the connection between the description of the total system at $t_i^e$ and that at $t_{i+1}$,
 we can temporarily employ the FR of $\R_I$, which is always possible,
 and transform the description to the FR of $\R_I$ at $t_i^e$.
 In the FR of $\R_I$, the state of the
 total system is also described by $|\Psi_{\mu^{(i)}}(t_i^e)\ra$ at $t_i^e$
 and has Schr\"{o}dinger evolution in the following times.

 At the time $t_{i+1}$ when the FR of $\R_{\{\mu^{(i+1)}\}}$ becomes valid,
 we can transform from the FR of $\R_I$ to the FR of $\R_{\{\mu^{(i+1)}\}}$.
 Suppose $U(t_{i+1},t_i^e)|\Psi_{\mu^{(i)}}(t_i^e)\ra$ satisfies Eq.~(\ref{cond-RS})
 for all values of $\mu^{(i+1)}$.
 Then, according to Postulate IV, with the probability
 \be \label{pii+1} p_{i+1,i} = \frac{\la \Psi_{\mu^{(i)}}(t_i^e) |U^\dag (t_{i+1},t_i^e)
 \PP_{\mu^{(i+1)}}U(t_{i+1},t_i^e)|\Psi_{\mu^{(i)}}(t_i^e)\ra}
 {\la \Psi_{\mu^{(i)}}(t_i^e)\ra  |\Psi_{\mu^{(i)}}(t_i^e)\ra  }, \ee
 the total system is described by
 \be |\Psi_{\mu^{(i+1)}}(t_{i+1})\ra = L_{\mu^{(i+1)}\mu^{(i)}}(t_{i+1},t_i^e)
 |\Psi_{\mu^{(i)}}(t_i^e)\ra  \label{PLi}\ee
 in the FR of $\R_{\{\mu^{(i+1)}\}}$,
 where we have introduced a transition operator $L_{\mu^{(i+1)}\mu^{(i)}}(t_{i+1},t_i^e)$,
 defined by
 \be L_{\mu^{(i+1)}\mu^{(i)}}(t_{i+1},t_i^e)
 \equiv \PP_{\mu^{(i+1)}} U(t_{i+1},t_i^e) \PP_{\mu^{(i)}}. \label{L} \ee
 For brevity, we may write $L_{\mu^{(i+1)}\mu^{(i)}}(t_{i+1},t_i^e)$ as $L_{i+1,i}$.

 To summarize, we have the following picture of evolution of the state of $\R+\E$
 in the FRs of $\R_{\{\mu^{(i)}\}}$,
 \bey \nonumber |\Psi_{\mu^{(0)}}(t_0)\ra   \xrightarrow[]{ \ U \ } |\Psi_{\mu^{(0)}}(t_0^e)\ra
 \xrightarrow[]{ \ L_{1,0} \ } |\Psi_{\mu^{(1)}}(t_1)\ra \xrightarrow[]{ \ U \ }
 \cdots \hspace{0.6cm}
 \\ |\Psi_{\mu^{(i)}}(t_i)\ra  \xrightarrow[]{ \ U\ } |\Psi_{\mu^{(i)}}(t_i^e)\ra
 \xrightarrow[]{ \ L_{i+1,i} \ }
 |\Psi_{\mu^{(i+1)}}(t_{i+1})\ra \cdots , \ \ \label{seq-UL}  \eey
 with probabilities for the jumps $L_{i+1,i}$ given in Eq.~(\ref{pii+1}).
 Generalization of the above description to the case of density operator is straightforward.


 \subsection{Measurement}
 \label{sect-measurement}

 In this section, making use of results given above, we discuss some properties of
 measurement. Specifically, we discuss a typical measurement process and derive the content
 of the axiom of measurement in the standard formalism of quantum mechanics in
 Sec.~\ref{sect-meas-axiom}, then, give a discussion for the phenomenon called
 collapse of state vector in Sec.~\ref{sect-collapse}.

 \subsubsection{Derivation of the content
 of the axiom of measurement in the standard formalism of quantum mechanics}
 \label{sect-meas-axiom}

 In a typical measurement process, the system $\cs$ to be measured is prepared, then enters into the
 interaction region of a RS $\R$ that can act as a measuring apparatus.
 After the interaction with $\R$, $\cs$ leaves some record as the value of a reference property
 of $\R$.
 The system $\cs$ is a part of the environment $\E$ of $\R$, but,
 for brevity, in this section,
 we do not explicitly mention the existence of other systems in the environment $\E$.

 Let us consider a state of $\R+\cs$, which is described by a normalized
 vector $ |\Psi_{\mu^{(0)}}(t_0)\ra $ at a time $t_0$ in the FR of $\R_{\{\mu^{(0)}\}}$.
 Beyond $t_0$, due to interaction between $\R$ and $\cs$, the FR of $\R_{\{\mu^{(0)}\}}$
 becomes invalid.
 Suppose at a later time $t = t_1$ Schr\"{o}dinger evolution of $ |\Psi_{\mu^{(0)}}(t_0)\ra $
 in the FR of $\R_I$, $|\Psi(t)\ra =U(t,t_0) |\Psi_{\mu^{(0)}}(t_0)\ra $,
 satisfies Eq.~(\ref{cond-RS}) for all the values $\mu$ of a division $\{\Pm\}$.
 Then, according to Postulate IV, the FR of $\R_{\{\mu\}}$ is valid at $t_1$
 and, as discussed in the previous section,
 in this FR the total system is described by one of the following vectors
 \be |\Psi_{\mu}(t_1)\ra =  {L_{\mu\mu^{(0)}}(t_1,t_0) |\Psi_{\mu^{(0)}}(t_0)\ra},
 \label{transi} \ee
 each with the probability $\| L_{\mu\mu^{(0)}}(t_1,t_0) |\Psi_{\mu^{(0)}}(t_0)\ra \|^2$,
 where $L$ is the transition operator defined in Eq.~(\ref{L}).

 For a measurement to be done, it should be possible to get information about the system
 $\cs$ from the value $\mu$ of the reference property $\{\mu\}$.
 This can be done, e.g., if $|\Psi_{\mu}(t_1)\ra$ is a product state of $\cs$ and $\R$,
 namely, $|\Psi_{\mu}(t_1)\ra = |s_\mu\ra |R_{ \mu}\ra$, where $|s_\mu\ra$ and $ |R_{\mu}\ra $
 are vectors in the Hilbert spaces of $\cs$ and $\R$, respectively.
 In this case, the state of $\cs $ is $|s_\mu\ra $ at $t_1$ in the FR of $\R_{\{\mu\}}$.

 Making use of the above results, for appropriately designed measurement schemes, we can derive
 the content of the axiom of measurement in the standard formalism of quantum mechanics.
 For this purpose, let us consider a measurement for an observable $A$ of the system $\cs$, which has
 normalized eigenvectors $|a\ra $ with eigenvalues $a$, $ A |a\ra =a|a\ra $.
 Suppose the initial vector has a product form,
 \be |\Psi_{\mu^{(0)}}(t_0)\ra= \left ( \sum_a c_a |a\ra \right ) |R_0 \ra . \label{ini-s} \ee
 The measurement may be designed in a way similar to that discussed in
 von Neumann's measurement theory \cite{Neumann-qm},
 such that the state of $\cs +\R$ in the FR of $\R_I$ has the following form
 after the interaction period of time,
 \be |\Psi (t_1)\ra = \sum_a c_a |a\ra |R_{\mu (a)}\ra , \label{mpr} \ee
 where $|R_{\mu (a)}\ra$ are normalized vectors in subspaces $\HH_{\R\mu}$, respectively,
 and $\mu$ is a function of $a$ satisfying $\mu(a) \ne \mu(a')$ for $a \ne a'$.
 Note that for this type of interaction process,
 the states $|a\ra $ are robust under the interaction between $\cs $ and $\R$
 and decoherence has taken place, since $\la R_{\mu (a)} | R_{\mu (a')} \ra =0$ for $a \ne a'$.

 If at the time $t_1$ the FR of $\R_{\{\mu\}}$ is valid as discussed above,
 then,  in the FR of $\R_{\{\mu\}}$, with the probability $|c_a|^2$,
 the state of $\cs+\R$ is described by
 \be |\Psi_{\mu(a)}\ra = |a\ra |R_{\mu (a)}\ra  \label{mea-a} \ee
 and the state of $\cs $ is the eigenstate $|a\ra $.
 This is just the content of the axiom of measurement in the standard formalism of
 quantum mechanics.
 Meanwhile, the state of $\R$ is $|R_{\mu (a)}\ra$
 and the function $\mu(a)$ implies that
 the value of $a$ can be inferred from the value $\mu$ of the reference property of
 $\R$ after the measurement.


 \subsubsection{A discussion of collapse of state vector}
 \label{sect-collapse}

 In the standard formalism of quantum mechanics, with the measuring apparatus outside
 its description, the change of the state of a measured system $\cs$ in a measuring process,
 e.g., from $\sum_a c_a |a\ra$ to $|a\ra $, is called a collapse of state vector.
 Making use of results obtained in the preceding section for measuring processes,
 here we give a further discussion for this phenomenon.

 As already stressed, when describing the total system, it is important to fix the employed FR.
 Let us first discuss the description given in the FR of $\R_I$.
 In this FR, the description changes in two steps.
 The first step, from $|\Psi_{\mu^{(0)}}(t_0)\ra$ in Eq.~(\ref{ini-s}) to
 $|\Psi (t_1)\ra$ in Eq.~(\ref{mpr}), is a unitary process,
 usually called a pre-measurement process.
 In the second step, the pure vector description $|\Psi (t_1)\ra$ is changed to a density
 operator description $\rho(t_1) =\sum_a |c_a|^2 |\Psi_{\mu(a)}\ra \la \Psi_{\mu(a)}|$.
 The second step is not a dynamical process, but, is a consequence of
 a requirement of the principle of
 consistent description, namely, $|\Psi(t_1)\ra $ and $\rho(t_1)$ being physically equivalent.
 Thus, there is no real process of collapse in the FR of $\R_I$, but a change
 between two physically equivalent descriptions.

 We may also describe the measuring process in non-trivial FRs of $\R$.
 Suppose we are only interested in descriptions given in a FR of
 $\R_{\{\mu^{(0)}\}}$ and in a FR of $\R_{\{\mu\}}$.
 Let us consider a situation discussed in the previous Sec.~\ref{sect-meas-axiom}.
 That is, the non-trivial FR of $\R_{\{\mu^{(0)}\}}$ is valid at $t_0$
 and the non-trivial FR of $\R_{\{\mu\}}$ is valid at a later time $t_1$,
 but, none of the two FRs is valid for $t\in (t_0,t_1)$ due to interaction between $\R$ and $\cs$.
 In this situation, as discussed in Sec.~\ref{sect-meas-axiom},
 there is a jump of the description of $\R+\cs$,
 from $|\Psi_{\mu^{(0)}}(t_0)\ra$ in Eq.~(\ref{ini-s}) at $t_0$
 in the FR of $\R_{\{\mu^{(0)}\}}$  to $|\Psi_{\mu(a)}\ra$ in Eq.~(\ref{mea-a}) at $t_1$
 in the FR of $\R_{\{\mu\}}$.
 The jump is described by the transition operator $L_{\mu\mu^{(0)}}(t_1,t_0)$
 in Eq.~(\ref{transi}).
 This jump from $|\Psi_{\mu^{(0)}}(t_0)\ra$ to $|\Psi_{\mu(a)}\ra$ is similar to a collapse
 of state vector, however, it does not happen instantly, but takes a time interval $(t_0,t_1)$.


 We remark that there exists no FR (within the scope of the theory), in which the process
 $|\Psi_{\mu^{(0)}}(t_0)\ra \to |\Psi_{\mu(a)}\ra$ can be described in a continuous way.
 For example, in the FR of $\R_I$, the description is continuous, however,
 the final description is not $|\Psi_{\mu(a)}\ra$, but the density operator $\rho(t_1)$.
 Meanwhile, the FRs of $\R_{\{\mu^{(0)}\}}$ and $\R_{\{\mu\}}$ are not valid
 between $t_0$ and $t_1$ due to the interaction between $\R$ and $\cs$.
 Within the scope of the theory, since details of the process $|\Psi_{\mu^{(0)}}(t_0)\ra \to
 |\Psi_{\mu(a)}\ra$ can not be described, the theory tells nothing about the dynamics
 of this transition.
 Whether the process may be described in more detail in some other theory, e.g.,
 when some other type of RS and FR are considered, needs
 future investigation.

 \subsection{Irreversibility}

 As discussed in Sec.~\ref{sect-allowed-H},
 the principle of consistent description imposes a restriction in
 physically allowed vectors in the total Hilbert space.
 Specifically, for a set $W_d$ including at least one non-trivial division,
 the allowed $\HH$-region of $W_d$ is usually smaller than the total Hilbert space.
 This restriction in vectors has an important consequence, i.e.,
 it breaks the time reversal symmetry:
 Although Schr\"{o}dinger equation has the time reversal symmetry, the time reversal
 of a physically allowed vector may lie in a region outside the allowed $\HH$-region.

 The irreversibility in the time evolution can be shown quantitatively
 by von Neumann entropy $S$ in the FR of $\R_I$,
 \be \label{vN-entropy} S = - {\rm Tr} \{ \rho \ln \rho \}. \ee
 For an initial normalized vector $|\Psi(t_0)\ra $, Schr\"{o}dinger evolution of the
 vector gives constant von Neumann entropy, $S(\Psi,t)=0$, where the dependence of
 $S$ on the state $|\Psi\ra $ and time $t$ is written explicitly.

 However, as discussed in Sec.~\ref{sect-FR-I}, $|\Psi(t_0)\ra $ may split into
 branches and $|\Psi(t)\ra $ can be physically equivalent a
 density-operator description $\rho_\Upsilon(t)$ in Eq.~(\ref{rho-fi}).
 The von Neumann entropy of $|\Psi(t)\ra $ is different from that of $\rho_\Upsilon(t)$.
 For  $\rho_\Upsilon(t)$,
 substituting Eqs.~(\ref{rho-fi}) and (\ref{Pan}) into Eq.~(\ref{vN-entropy}), we have
 the following expression of von Neumann entropy, with the dependence on $\Upsilon$
 written explicitly,
 \be \label{vN-St} S(\Upsilon,t) = - \sum_\alpha P_\alpha(t) \ln P_{\alpha}(t) . \ee
 When no splitting happens, $S(\Upsilon,t)$ keeps constant
 as a result of the unitary evolution.
 However, at each splitting time $t_i$ along the paths of $\Upsilon$,
 the entropy $S(\Upsilon,t)$ obtains a discontinuous increment.
 Hence,  for each physically allowed vector $|\Psi(t_0)\ra $, $S(\Upsilon,t)$ either
 keeps a constant or increases with time, but never decreases.

 The above discussions show that a state of the total system usually
 does not have a unique value of von Neumann entropy $S$.
 For a pure vector description,  $S(\Psi,t)=0$.
 On the other hand, when some non-trivial reference property becomes valid at a time $t_i$,
 there can be a discontinuous change in the density
 operator description $\rho_\Upsilon(t)$ and an increment in the entropy $S_N(\Upsilon,t)$.
 These results suggest that von Neumann entropy may be given an information interpretation:
 Constant von Neumann entropy means that no new information is obtainable by measurement,
 while an increment of the entropy means that new information is available by taking
 the value(s) of some reference property of $\R$.

 As discussed in Sec.~\ref{sect-FR-I}, there is some arbitrariness in choosing
 $\Upsilon$, e.g., in choosing the splitting times.
 This arbitrariness is related to the freedom for an observer to determine the times of
 taking measurement and the reference properties to be recorded.
 The upper bound of $S(\Upsilon,t)$ for all possible trees under a given $W_d$
 does not have this arbitrariness,
 hence, we may introduce
 \be S_m(t) = \sup_\Upsilon S(\Upsilon,t) . \ee
 It is the maximum entropy, corresponding to a tree composed of paths that split whenever possible.
 The quantity $S_m(t)$ also either increases or keeps a constant with increasing time,
 indicating the irreversibility of time evolution.

 There is an old problem of irreversibility of macroscopic systems,
 as stated in the second law of thermodynamics.
 This problem lies in the heart of the foundation of statistical physics
 and has been a topic of debating for more than one hundred years.
 One idea regarding this problem is that the phenomenon of thermodynamic irreversibility
 may has its origin in the allowed initial condition, that is, only certain type of
 initial condition should be considered \cite{Zeh99}.
 The above discussed irreversibility due to the restriction in allowed $\HH$-region
 may shed new light in this old problem of
 irreversibility.
 A difference between the two cases lies in the fact that here the restriction in physically
 allowed vectors comes from the usage of a system as a RS, while the role of RS is not
 emphasized in thermodynamic irreversibility.
 To bridge the two situations is not an easy task and
 we would leave it to future investigation.

 \section{A model: RS as a two-level system}
 \label{sect-model}

 In this section we use a simple model to illustrate several points of the
 theory proposed above.
 In particular, we discuss a decoherence mechanism for Eq.~(\ref{cond-RS}) to hold
 with an arbitrarily small $\epsilon_R$ and show that a class of vector belongs to the
 allowed $\HH$-region of an employed set $W_d$ of divisions.
 We also discuss relation between $\tau_s$ and decoherence time.

 In this model, the RS is a two level system with energy eigenstates $|k\ra$,
 $H_\R|k\ra =E_k|k\ra $ for $k=1,2$.
 The system has only two divisions that satisfy the stability condition (\ref{sta}),
 namely, the trivial division $\{I\}$ and the division $\{\PP_k\}=\{ |1\ra \la 1| ,|2\ra \la 2|\}$.
 We consider a set $W_d$ composed of these two divisions, $W_d =\{ \{I\}, \{\PP_k\} \}$.
 It is seen that $\ov \PP_1=\PP_2$ and $\ov \PP_2=\PP_1$.

 For the simplicity  in discussion, we assume that Eq.~(\ref{cond-RS}) is required
 to be satisfied for both $k=1$ and 2 of the property $\{ k \}$.
 In this case, the condition (\ref{cond-RS-ov}) gives
 \be \PP_2 H_I \PP_1 |\Psi\ra \doteq 0 \ \ \text{and} \ \
 \PP_1 H_I \PP_2 |\Psi\ra \doteq 0, \label{cond-RS-2} \ee
 hence, for a $|\Psi\ra $ satisfying this special case of Eq.~(\ref{cond-RS}),
 the interaction Hamiltonian $H_I$ has effectively the following simple form,
 \be \ww H_I \doteq |1\ra \la 1| H^{I\E}_{1} + |2\ra \la 2| H^{I\E}_{2}, \label{wwHI} \ee
 where $H^{I\E}_{k}$ of $k=1,2$ are Hermitian operators in the Hilbert space of $\E$.

 \subsection{An expression of $\D_{\alpha\alpha'}^m$ for $t \in [t_1,t_1^e]$}

 As in Sec.~\ref{sect-fix-FR}, we assume that the condition (\ref{cond-RS}) is
 satisfied in the time intervals $[t_0,t_0^e]$,
 $[t_1,t_1^e]$, $\ldots$, and is not satisfied in the intervals $(t_0^e,t_1)$,
 $(t_1^e,t_2)$, etc..
 For the simplicity in discussion, we assume that these time scales, $t_0^e$, etc.,
 are path-independent.
 The initial normalized vector is written as
 $ |\Psi(t_0)\ra  =\sum_{m^{(0)}=1}^2 |m^{(0)}\ra |\varphi_{m^{(0)}}(t_0)\ra $,
 where
 \be |\varphi_{m^{(0)}}(t_0)\ra = \la m^{(0)}|\Psi(t_0)\ra  . \ee
 Here, $m^{(i)}= 1^{(i)},2^{(i)}$ correspond to $k=1,2$, respectively,
 for $i=0,1,\ldots $.

 For times $t\in [t_0,t_0^e]$, $H_I$ has the form of $\ww H_I$ in Eq.~(\ref{wwHI}) and
 \be |\Psi(t)\ra \doteq \sum_{m^{(0)}=1}^2 |\Psi_{m^{(0)}}(t)\ra ,  \ee
 where
 \bey |\Psi_{m^{(0)}}(t)\ra = \sum_{m^{(0)}=1}^2 e^{-iE_{m^{(0)}}(t-t_0)/\hbar}|m^{(0)}\ra
 |\varphi_{m^{(0)}}(t)\ra , \eey
 with
 \bey |\varphi_{m^{(0)}}(t)\ra  = U^\E_{m^{(0)}}(t,t_0)  |\varphi_{m^{(0)}}(t_0)\ra . \eey
 Here $U^\E_{m^{(0)}}(t,t_0)$ is defined by
 \be U^\E_{m^{(i)}}(t,t_i) = \exp \left \{ -\frac i{\hbar} \left ( H^{I\E}_{m^{(i)}}+H_\E \right )
 (t-t_i) \right \} \ee
 with $i=0$ and is an unitary operator in the Hilbert space of $\E$,
 given by the effective Hamiltonian
 $H^{\E{\rm eff}}_{m^{(0)}} = H^{I\E}_{m^{(0)}}+H_\E$.
 Since $ |\Psi_{m^{(0)}}(t)\ra$ lies in the subspace $\PP_{m^{(0)}}$,
 Eq.~(\ref{D-diag}) is satisfied for $t\in [t_0,t_0^e]$.


 For the time interval $(t_0^e,t_1)$, $H_I$ can not be approximated by $\ww H_I$ and
 we have to write the full expression,
 \be |\Psi_{m^{(0)}}(t)\ra = U(t,t_0^e) |\Psi_{m^{(0)}}(t_0^e)\ra ,
 \  \ t \in (t_0^e,t_1). \ee
 At $t_1$, $|\Psi(t)\ra $ satisfies the condition (\ref{cond-RS}) again
 and, as discussed in Sec.~\ref{sect-FR-I},
 we can let each branch $|\Psi_{m^{(0)}}(t)\ra$ split into two sub-branches.

 For $t\in [t_1, t_1^e]$, direct derivation shows
 \be |\Psi (t)\ra \doteq \sum_{m^{(0)}=1}^2 \sum_{m^{(1)}=1}^2 |\Psi_{m^{(0)}m^{(1)}}(t)\ra , \ee
 where
 \be |\Psi_{m^{(0)}m^{(1)}}(t)\ra = e^{-i\Theta_{m^{(0)}m^{(1)}}(t)}|m^{(1)}\ra
 |\varphi_{m^{(0)}m^{(1)}}(t)\ra . \ \ee
 Here
 \bey \Theta_{m^{(0)}m^{(1)}}(t)= \frac 1{\hbar} \left [ E_{m^{(1)}}(t-t_1) + E_{m^{(0)}}(t_0^e-t_0)
 \right ]
 \\ \nonumber |\varphi_{m^{(0)}m^{(1)}}(t)\ra = U^\E_{m^{(1)}}(t,t_1) U_{m^{(1)}m^{(0)}}(t_1,t_0^e)
 \\ \cdot  U^\E_{m^{(0)}}(t_0^e,t_0) |\varphi_{m^{(0)}}(t_0)\ra , \label{phi-second}
 \eey
 where
 \be U_{m^{(1)}m^{(0)}}(t_1,t_0^e) = \la m^{(1)}|U(t_1,t_0^e)|m^{(0)}\ra \label{Um10} \ee
 is an operator in the Hilbert space of $\E$.
 Then, the quantity $\D_{\alpha \alpha'}^m$ can be written as follows,
 \bey \nonumber \D_{\alpha \alpha'}^m = \la\Psi_{m^{(0)}m^{(1)}}(t)|
 \PP_m |\Psi_{{m'}^{(0)}{m'}^{(1)}}(t)\ra
 \\ =  \delta_{{m'}^{(1)}m} \delta_{m^{(1)}m } e^{-i\Delta \Theta}  D_\varphi, \label{Daa-model}
 \eey
 where
 \bey D_\varphi = \la \varphi_{m^{(0)}m }(t)|\varphi_{{m'}^{(0)}m}(t)\ra .
 \eey
 Making use of Eq.~(\ref{phi-second}), we have the following explicit expression for $D_\varphi$,
 \bey \nonumber D_\varphi = \la \varphi_{m^{(0)}}(t_0)| U^{\E \dag}_{m^{(0)}}(t_0^e,t_0)
 U^\dag_{mm^{(0)}}(t_1,t_0^e) \hspace{1cm}
 \\ \hspace{1cm} U_{m{m'}^{(0)}}(t_1,t_0^e) U^\E_{{m'}^{(0)}}(t_0^e,t_0) |\varphi_{{m'}^{(0)}}(t_0)
 \ra . \label{D-varphi}
 \eey

 \subsection{Conditions for decoherence}
 \label{cond-decoh}

 The value of $D_\varphi$ is clearly initial-state dependent.
 Here, we consider a special class of initial vector, which has a product form
 $|\Psi(t_0)\ra = |R\ra |\varphi_0\ra $.
 In this case, $|\varphi_{{m'}^{(0)}}(t_0)\ra = |\varphi_{{m}^{(0)}}(t_0)\ra =|\varphi_0\ra$.
 To estimate $D_\varphi$, we note that it can be obtained by inserting
 $U^\dag_{mm^{(0)}} U_{m{m'}^{(0)}}$ into the following quantity,
 \bey  M_D = \la \varphi_0| U^{\E \dag}_{m^{(0)}}(t_0^e,t_0) \  {}^{\downarrow} \
 U^\E_{{m'}^{(0)}}(t_0^e,t_0) |\varphi_0\ra . \ \ \label{D-phi-app}
 \eey
 The quantity $M_D$ is the so-called Peres fidelity \cite{Peres84} or
 quantum Loschmidt echo.
 Generally, this fidelity is defined as the overlap of the time evolution of the same
 initial state under two slightly different Hamiltonians,
 \be M(t) =| \la \Psi_0|{\rm exp}(iHt/ \hbar ) {\rm exp}(-iH_0t / \hbar)
 |\Psi_0 \ra |^{2}, \label{mat} \ee
 where $H_0$ and $H$ are the unperturbed and perturbed Hamiltonians, respectively,
 $ H=H_0 + \epsilon V $,  with $\epsilon $ a small quantity and $V$ a generic perturbation.

 The Peres fidelity is a measure of the stability of quantum motion under small perturbation
 and has been found useful in the study of decoherence \cite{GPSS04,WGCL08}.
 It has been found to have a fast decay in quantum irregular systems with sufficiently
 large Hilbert space: Gaussian decay for sufficiently weak perturbation $\epsilon V$
 and an exponential decay for not very weak perturbation \cite{LE-rev,wwg-LEc}.

 For $M_D$ in Eq.~(\ref{D-phi-app}), the perturbation in the definition of the
 Peres fidelity  takes the form
 \be \epsilon V = H^{I\E}_{m^{(0)}} - H^{I\E}_{{m'}^{(0)}} . \ee
 For a sufficiently large and irregular environment $\E$ and not extremely weak perturbation
 $eV$, $M_D$ has typically an exponential decay,
 \be M_D \sim e^{-(t_0^e-t_0) / \tau_d} \ \  \ \ \text{for} \ m^{(0)} \ne {m'}^{(0)}. \label{MD-decay}
 \ee
 Here $\tau_d$ is a decoherence time, which is determined by both the perturbation
 $\epsilon V$ and the effective Hamiltonian $H^{\E{\rm eff}}_{m^{(0)}}$.

 When the time period $(t_0,t_0^e)$ is sufficiently long,  $t_0^e-t_0 \gg \tau_d$,
 $M_D$ in Eq.~(\ref{MD-decay}) is negligibly small.
 This implies that the vector $U^{\E}_{m^{(0)}}(t_0^e,t_0)|\varphi_0\ra$ is usually
 ``far'' from $ U^\E_{{m'}^{(0)}}(t_0^e,t_0) |\varphi_0\ra$.
 A reasonable assumption, which may hold in most cases, is that operators like
 $U_{m{m'}^{(0)}}(t_1,t_0^e)$
 defined in Eq.~(\ref{Um10}) does not bring together two such ``far'' vectors in $\HH_\E$.
 Thus, $D_\varphi$ in Eq.~(\ref{D-varphi}) is also small.
 Therefore, as long as $(t_0,t_0^e)$ is sufficiently long,
 for whatever small but finite $\epsilon_R$, we have
 \bey \D_{\alpha \alpha'}^m  \doteq \delta_{{m'}^{(1)}m^{(1)}}
 \delta_{{m'}^{(0)}m^{(0)}} \D_{\alpha \alpha}^m  ,
 \eey
 i.e., Eq.~(\ref{D-diag}) is satisfied.
 If we require that $\tau_s \gg \tau_d$, then,  the
 stability of reference property can guarantee the validity of the principle of
 consistent description.

 It is straightforward to generalize of the above discussions to times beyond $t_1^e$,
 $t_2$, etc..
 Finally, we reach the following conclusion.
 The principle of consistent description [Eq.~(\ref{D-diag})] is satisfied at least for
 most of initial vectors $|\Psi(t_0)\ra $ that have the product form,
 when the following conditions are satisfied:
 (1) The environment is sufficiently large and irregular,
 (2) there is enough difference between $H_1^{I\E}$ and $H_2^{I\E}$,
 and (3) $(t_i^e-t_i) \gg \tau_d$ for all times $t_i^e<t$.
 We remark that interaction in the time intervals $(t_i^e,t_{i+1})$
 may induce decoherence as well; we do not discuss it here, because of the
 mathematical difficulty of the topic.

 \section{Comparison with some interpretations of quantum mechanics}
 \label{sect-ot-int}

 In this section, we compare the theory proposed in this paper with
 some interpretations of quantum mechanics.

 \subsection{Many-worlds interpretations}


 Many-worlds interpretations of quantum mechanics \cite{Everett57,Deutsch85}, denoted by
 MWI in what follows, have two main assumptions.
 Namely, (i) Schr\"{o}dinger equation holds universally,
 and (ii) the state vector of the total system splits constantly into branches.

 In the theory here, Schr\"{o}dinger equation holds universally in each valid FR of $\R$,
 though a FR (except the FR of $\R_I$) may become invalid at some time.
 As discussed in Sec.~\ref{sect-FR-I} and illustrated in Fig.~\ref{fig-bra},
 in the FR of $\R_I$, the unitary evolution of an initial vector can be
 physically equivalent to a description in which the vector splits
 at some times into branches.
 This branching picture of evolution is similar to that given in MWI.

 Despite the above mentioned similarities, the theory here has conceptual and fundamental
 differences from MWI.
 (i) In the theory here, the description of the world may split into branches,
 but, there exists only one real world,
 unlike in some of the MWI  which allows the existence of many worlds at the same time.
 (ii) In the theory here, only vectors in some allowed region in the Hilbert space
 can be associated with physical states, while there is no such a restriction in MWI.
 (iii) The theory here gives an explicit condition for a splitting to happen,
 which is not given explicitly in MWI.

 \subsection{Consistent-histories interpretations}

 In consistent-histories interpretations of quantum mechanics, denoted by CHI in the following,
 the time evolution of a quantum system has a stochastic nature and is described by
 (quantum) consistent histories
 \cite{Griffiths84,Grif96,Griff02,Omnes88,Omnes92,Omnes99,GH90,GH93}).
 Each consistent history is composed of a sequence of events
 represented by time-ordered projection operators,
 with unitary connection between each two successive events,
 and has to satisfy certain consistency condition.
 The set of histories given from the same set of complete orthogonal projection operators
 is called a family.
 A basic rule in CHI is the so-called single family (framework) rule,
 stating that a meaningful description must employ a single consistent family.

 There exist some formal similarity between expressions given in CHI and some
 expressions given in the theory here.
 In fact, substituting Eq.~(\ref{Psi-alpha}) into Eq.~(\ref{Da}),
 it is seen that the quantity $\D^\mu_{\alpha\alpha'}$ can be written in a form with
 formal similarity to the so-called decoherence functional ${\cal{D}}(\beta , \beta ')$
 in CHI \cite{GH90,GH93},
 \bey \nonumber {\cal{D}}(\beta,\beta') = \tr \left [ P^{(n)}_{\beta_n} U(t_n,t_{n-1})
    \ldots P^{(1)}_{\beta_1} U(t_{1},t_{0}) \rho(t_0)   \right .
 \\  \left . U^{\dag}(t_{1},t_{0}) P^{(1)}_{\beta_1'}
  \cdots  U^{\dag}(t_n,t_{n-1}) P^{(n)}_{\beta_n'}  \right ] , \ \
 \label{Da-ch} \eey
 where $\beta $ indicates a history and $ P^{(j)}_{\beta_j}$
 of $j=1, \ldots , n$ denote projection operators in the history $\beta$.
 The summation $\sum_\mu \D^\mu_{\alpha\alpha}$ gives the probability for the realization of
 a path $\alpha$, correspondingly,
 ${\cal{D}}(\beta,\beta)$ gives the probability assigned to a consistent history $\beta$ in CHI.
 Equation (\ref{D-diag}) has the same formal form as the consistency condition in CHI,
 \be \label{CHI-con} {\cal{D}}(\beta , \beta ') =\delta_{\beta \beta'} {\cal{D}}(\beta , \beta ).
 \ee

 However, the difference between the theory here and CHI is more profound.
 (i) Unlike in CHI,
 dynamical description of the time evolution of the total system is allowed in the theory here,
 in particular, universal Schr\"{o}dinger evolution in the FR of $\R_I$.
 (ii) Equations (\ref{D-diag}) and (\ref{CHI-con}) have different physical meanings:
 Equation (\ref{D-diag}) imposes a restriction to physically allowed initial vectors in the
 theory here.
 While in CHI Eq.~(\ref{CHI-con}) selects projection operators that can form
 consistent histories, without any restriction to initial vectors.
 (iii) The projection operators $\Pm$ here are related to properties of the RS,
 while projection operators
 in CHI are not definitely related to any part of the total system.

 \section{Summary and discussions}
 \label{sect-conclusion}

 In this paper, we have proposed a quantum theory for a total system composed of a
 reference system (RS) and an environment.
 The theory is based on four basic postulates, which have, loosely speaking,
 the following contents.
 (i) A FR is related to a reference property and a reference property of a RS
 has a definite value and is sufficiently stable in its own FR.
 (ii) States of the total system are associated with vectors in the total Hilbert space.
 (iii) Schr\"{o}dinger equation is the dynamical law in each valid FR.
 (iv) Satisfying certain condition, a property of a system can be regarded as
 a reference property with certain probability.
 The four postulates lead to multiple descriptions for the same state of the
 total system and it is necessary to introduce a principle for consistent description,
 which guarantees the consistency of multiple descriptions,
 stating  that descriptions given in the same FR
 for the same state must be physically equivalent.
 Here, physical equivalence means being experimentally indistinguishable.

 The most significant consequence of the four postulates and the principle of consistent
 description is a restriction in physically allowed vectors in the Hilbert space.
 As a result, time evolution can be irreversible for a set $W_d$
 including at least one non-trivial division.
 This irreversibility may shed new light in the old problem of irreversibility
 stated in the second law of thermodynamics.

 \acknowledgments

 The author is grateful to Yan Gu, A.J.~Leggett, Hong Zhao, Jiangbin Gong and
 J.~Garcia for valuable discussions and suggestions.
  This work is partially supported by Natural Science Foundation of China Grant
No.~10775123 and the start-up funding of USTC.

\appendix


\begin{thebibliography}{99}
\bibitem{note-observer} In many theories, in particular, in the theory of relativity,
 observer is mentined explicitly, emphasizing on the fact that measurement is performed
 and description is made by the observer.
 The purpose of the theory proposed here is to give a description for measuring apparatus,
 as well as for a RS, in the language of quantum mechanics.
 Here, we do not need to mention the observer explicitly.
 Besides, we do not want to give any implication that the consciousness of the observer may play
 any role in the theory here.
\bibitem{note-RF} The concepts of FR and RS here are different from the concept of reference
 frame (reference system) used in the field of quantum information
 (see, e.g., Refs.~\cite{BRS07,PY07} and references therein),
 where a reference frame is referred to a system with
 respcet to which unspeakable information is defined.
\bibitem{note-bohr} This triple roles of the RS discussed in the theory proposed here
 has some similarity to Bohr's
 idea that an observer may play both the role of an ``actor'' and the role of a ``spectator''.
 A major difference is that in the theory here it is the RS, not the observer,
 that plays the essential role.
\end{thebibliography}
 \end{document}